\documentclass[fleqn,usenatbib,useAMS]{mnras}

\DeclareRobustCommand{\VAN}[3]{#2}
\let\VANthebibliography\thebibliography
\def\thebibliography{\DeclareRobustCommand{\VAN}[3]{##3}\VANthebibliography}

\usepackage{graphicx}
\usepackage{amsmath}
\usepackage{amssymb}

\usepackage[varg]{txfonts}
\usepackage[T1]{fontenc}

\newcommand{\bmx}{\bmath{x}}
\newcommand{\bmv}{\bmath{\varv}}

\DeclareMathOperator{\sech}{sech}
\defcitealias{An2021}{Paper~I}

\usepackage{xcolor}
\newcommand{\hl}[1]{\textcolor{magenta}{#1}}
\renewcommand{\hl}[1]{#1}

\title[Charting galactic accelerations II]{Charting galactic accelerations II: how to `learn' accelerations in the solar neighbourhood}
\author[A.~P.~Naik et al.]
{A.~P.~Naik$^1$\thanks{E-mail: aneesh.naik@nottingham.ac.uk}, J.~An$^2$, C.~Burrage$^1$, and N.~W.~Evans$^3$ 
\\$^1$ School of Physics \& Astronomy, University of Nottingham, University Park, Nottingham NG7~2RD, UK;
\\$^2$ Center for Theoretical Astronomy, Korea Astronomy \& Space Science Institute, 776 Daedeok-daero, Yuseong-gu, Daejeon 34055, Korea (South);
\\$^3$ Institute of Astronomy, University of Cambridge, Madingley Rd., Cambridge CB3~0HA, UK.}
\date{Accepted XXX. Received YYY; in original form ZZZ}
\pubyear{2021}

\begin{document}
\label{firstpage}
\pagerange{\pageref{firstpage}--\pageref{lastpage}}
\maketitle


\begin{abstract}
Gravitational acceleration fields can be deduced from the collisionless Boltzmann equation, once the distribution function is known. This can be constructed via the method of normalizing flows from datasets of the positions and velocities of stars. Here, we consider application of this technique to the solar neighbourhood. We construct mock data from a linear superposition of multiple `quasi-isothermal' distribution functions, representing stellar populations in the equilibrium Milky Way disc. We show that given a mock dataset comprising a million stars within 1~kpc of the Sun, the underlying acceleration field can be measured with excellent, sub-percent level accuracy, \hl{even in the face of realistic errors and missing line-of-sight velocities.} The effects of disequilibrium can lead to bias in the inferred acceleration field. This can be diagnosed by the presence of a phase space spiral, which can be extracted simply and cleanly from the learned distribution function. We carry out a comparison with two other popular methods of finding the local acceleration field (Jeans analysis and 1D distribution function fitting). We show our method most accurately measures accelerations from a given mock dataset, particularly in the presence of disequilibria. 
\end{abstract}



\section{Introduction}
\label{S:Intro}

Given a map of the gravitational acceleration field within a kiloparsec of the Sun, we could learn a wealth of information about the current state of our Galaxy, the distribution of matter (both dark and luminous) and the nature of gravity. For example, if the acceleration due to the luminous component is known, then we can calculate the density distribution of dark matter, uncovering any substructures and measuring the ambient dark matter density in the solar system. This latter number is of great importance in particle physics, as it is a key parameter in the interpretation of results of dark matter direct detection experiments \citep{Read2014, deSalas2020}. Alternatively, we can use the direction of the acceleration vectors to constrain alternative theories of gravity such as Modified Newtonian Dynamics \citep[MOND;][]{Milgrom1983} or similar. In these theories, there is no dark matter, so the total acceleration is necessarily co-linear with the acceleration due to the baryons, even if the modifications to gravity alter its magnitude \citep{Loebman2014}.

Unfortunately, direct acceleration measurements are challenging. Even so, promising steps have been recently taken in this direction, employing measurements of pulsar orbital decay \citep{Chakrabarti2021}, \hl{which give the relative acceleration of a few pulsar systems with respect to the Solar system. These can then be converted into absolute accelerations using a measurement of the Solar System acceleration, thus providing a small number of direct samples of the Galaxy's acceleration field \citep[e.g.,][]{Bovy2020}.} Future high-precision radial velocity spectrographs promise greater statistical power still \citep{Silverwood2019, Chakrabarti2020}.

In the meantime, we must instead adopt an alternative approach: inferring accelerations (or equivalently the gravitational potential) statistically from the positions and velocities of the stars. If discrete stellar encounters are neglected, then the stellar distribution function (DF), i.e.\ the probability distribution of the stars in six-dimensional ($\bmx, \bmv$) phase space, can be related to gravitational accelerations via the collisionless Boltzmann equation (CBE),
\begin{equation}
\label{E:CBE}
    \frac{\partial f}{\partial t} + \bmath{\varv \cdot \nabla_x}f - \bmath{\nabla_\varv}f \bmath{\cdot\, \nabla_x}\!\Phi = 0,
\end{equation}
where $f$ is the DF and $\Phi$ is the gravitational potential.

It can be difficult to constrain the full DF with a statistically small dataset, so it is often preferable to work with the second moments of the DF, i.e.\ the velocity dispersions, which can be related to the acceleration field via the Jeans equations \citep[e.g.,][]{Hagen2018, Sivertsson2018, Guo2020, Salomon2020}. While velocity dispersions are comparatively easy to measure from kinematic data, we lose much of the information content of the data compared with techniques working directly with the DF. For this reason, many studies have instead adopted the latter approach, typically in one dimension \citep[e.g.,][]{Schutz2018, Buch2019, Widmark2019a, Widmark2019b, Widmark2021WeighingDisc, Li2021}. Note that either treatment necessitates the assumption of dynamical equilibrium, so that the time-derivative term in equation~(\ref{E:CBE}) can be neglected. The review articles by \citet{Read2014} and \citet{deSalas2020} give good overviews of how these methods work in practice.

Despite the statistical advantages of the latter class of technique, there are some limitations. Typical DF models are constructed under a series of assumptions, such as separability, isothermality, and various spatial symmetries. In the era of `big data', it is worth examining whether an alternative approach can do better justice to the full richness of contemporary datasets and their statistical power. In \citet[hereafter Paper I]{An2021}, we outlined just such a methodology. Inspired by an idea first proposed by \citet{Green2020}, we described a technique in which a non-parametric DF can be constructed directly from the data using modern deep learning techniques. Such an approach is highly flexible; the resulting learned DF is untrammelled by the limitations of an analytic model, and is instead free to capture the full richness of the training data. After learning a DF in this manner, \citetalias{An2021} shows how to convert the DF into an acceleration map, via an exact inversion of the CBE under the assumption of equilibrium. 

In \citetalias{An2021}, we provided a basic demonstration of our technique with a mock dataset representing a simple, spherical distribution of stars. In this article, we provide demonstration of the same technique in much more complex context: mock data on stellar kinematics in the solar neighbourhood. The reason to confine ourselves to mock data for the moment is to gain insights into the biases and limitations of the technique before we apply it to real data in a companion paper.

A major obstacle is that the assumption of dynamical equilibrium is not necessarily a good one. Various non-equilibrium structures have been observed in kinematics of the Milky Way (MW) disc stars, such as warping of the disc \citep{Schoenrich2018}, north-south asymmetries \citep{Salomon2020}, and the well-known phase spiral \citep{Antoja2018}. Incorrectly assuming that a stellar population is in equilibrium will lead to bias in resulting dynamical inferences (\citealt{Banik2017}; \citetalias{An2021}). To quantify this effect, we additionally examine the application of our methodology to a mock dataset resembling a perturbed Galactic disc.

This article is structured as follows. In the following section (Section~\ref{S:Methods}), we recapitulate the methodology described in \citetalias{An2021} -- namely, our algorithm to recover acceleration fields from 6D kinematics. After that, in Section~\ref{S:Data}, we describe the mock datasets we use to test this method. We generate stellar positions and velocities from realistic models of the MW disc. We calculate radial and vertical accelerations within $\sim 1~\mbox{kpc}$ of the Sun from a mock dataset in Section~\ref{S:Results}. Then, Section~\ref{S:Comparison} compares the accuracy of our measured accelerations against those produced using other techniques, specifically the Jeans analysis method of \citet{Salomon2020} and the 1D DF fitting method of \citet{Widmark2021WeighingDisc}. Finally, Section~\ref{S:Conclusions} provides a discussion and concluding remarks.

\section{Methods}
\label{S:Methods}

The two steps in our method are \citepalias[see][]{An2021}:
\begin{enumerate}
    \item Given a stellar kinematic dataset, we use a probability estimation technique to `learn' the underlying DF.
    \item From the learned DF, we then calculate the gravitational acceleration field using an inversion of the (time-independent) CBE.
\end{enumerate}

\subsection{Learning the DF}
\label{S:Methods:LearningDF}

Given data sampled from some unknown distribution, the problem of trying to derive the underlying probability distribution is known as `probability estimation'. In our case, the data are the positions $\bmx$ and velocities $\bmv$ of stars, and the probability we wish to estimate is the stellar DF $f(\bmx, \bmv)$, i.e.\ the probability density function of stars in phase space. One way to do this is to write down some parametric model for the probability density then compare the model's predictions with the data until the parameters are optimized. In essence, this is the technique employed by the majority of studies of stellar dynamics, whether they work directly with the DF or with its moments.

We instead adopt a different methodology: we estimate a \emph{non-parametric} DF directly from the data. This data-driven approach has the distinct advantage of being untrammelled by the limitations and underlying assumptions of an explicit model. While non-parametric probability estimation techniques have long existed (e.g., kernel density estimation), recent years have seen a surge of interest in machine learning techniques, which in turn has led to a proliferation of powerful probability estimation algorithms. We employ one such novel algorithm: `normalizing flows' \citep{Rezende2015}.

The idea behind normalizing flows is simple: we can generate a complex probability distribution by repeatedly transforming a simple one, such as a Gaussian. The input Gaussian can be said to `flow' through the series of transformations, and after each transformation a (multiplicative) normalizing factor is applied to the new probability distribution to ensure that it is properly normalized, hence the name `normalizing flows'. 

To see this, consider a continuous random variable $z$, with probability density function $p_z(z)$. We can define a new variable $x \equiv f(z)$, with the only requirement being that the function $f$ is bijective\footnote{Technically we also assume both $f$ and $f^{-1}$ are continuously differentiable; most practical applications are limited to such transformations.} (and thus invertible). The probability distribution for $x$ is then
\begin{equation}
    p_x(x) = p_z\!\left(f^{-1}(x)\right)\, \left\vert \det \left( \frac{\partial f^{-1}(x)}{\partial x} \right) \right\vert.
\end{equation}
The determinant on the right-hand side is the normalizing factor.

This can be generalized to a series of bijective transformations, i.e.\ $x \equiv f(z) = (f_K\circ f_{K-1}\circ \cdots \circ f_2\circ f_1) (z)$. Now, the probability distribution for $x$ is (here $f^{-1}=f_1^{-1}\circ f_2^{-1}\circ \cdots \circ f_{K-1}^{-1}\circ f_K^{-1}$)
\begin{equation}
    p_x(x) = p_z\!\left(f^{-1}(x)\right)\, \prod_{i=1}^{K} \left\vert \det \left( \frac{\partial f_i^{-1}(x)}{\partial x} \right) \right\vert.
\end{equation}
By suitably choosing the input distribution $p_z$ and the transformations $f_i$, we can generate arbitrarily complex probability distributions. In practice, most applications use a simple unit Gaussian for $p_z$. Furthermore, we typically give each of the transformations $f_i$ the same parametric form, although each transformation can take different parameters. We can then construct a suitable loss function, such that minimizing the loss function (with respect to the transformation parameters) corresponds to generating a probability distribution $p_x$ that best describes the distribution of $x$.

In practice, we constructs an \textit{ensemble} of estimators. Before training, each flow in the ensemble is initialized with different parameters. The high complexity of a typical loss landscape means that each flow is then likely to end its training at a different local minimum, i.e.\ each flow learns a slightly different probability distribution. The probability $p_x$ (and its gradients) is then given by the mean over all the estimators.

In summary, the user need only specify:
\begin{enumerate}
    \item the parametric form of an individual transformation,
    \item the loss function,
    \item the number of `units' (i.e.\ transformations),
    \item the number of flows in the ensemble.
\end{enumerate}

It is worth remarking that the target distribution, the learned DF, is guaranteed to be a well-behaved probability distribution, i.e.\ it is positive everywhere and has unit normalization. The latter requirement restricts the space of usable transformations to bijective functions, and this space is then restricted further by the desire for computational efficiency. Different normalizing flow techniques differ primarily in their choices of these transformations, as well as the base distributions and flow architectures.

More detailed descriptions of normalizing flows are given in the article by \citet{Rezende2015} first describing the algorithm, and the recent review articles by \citet{Kobyzev2020} or \citet{Papamakarios2021}. Also, \citetalias{An2021} provides a discussion of the advantages of normalizing flows over kernel density estimation.

We do not impose any further physicality requirements on the learned DF. For example, the acceleration vectors calculated from the DF (see Section~\ref{S:Methods:DFtoAcc}) could in principle show negative divergences (i.e.\ negative mass densities) or non-zero curls (i.e.\ non-conservative forces). This is advantageous as it allows us to probe deviations from Newtonian gravity or the effects of disequilibrium.

\citet{Green2020} employed a species of normalizing flow called `neural spline flows' \citep{Durkan2019}. Applied to our mock datasets, we find that neural spline flows struggle with hard edges of the sample volumes. We instead use `masked autoregressive flows' \citep[MAFs;][]{Papamakarios2017}. For each mock dataset, we train an ensemble of 20 estimators, each with 8 transformations along the flow, each transformation being an artifical neural network with one hidden layer of 64 units. We use the implementation of MAFs in the publicly available software package \textsc{nflows},\footnote{\textsc{nflows}: normalizing flows in \textsc{pytorch}, \doi{10.5281/zenodo.4296287}} and train the estimators using the gradient descent algorithm \textsc{adam} \citep{Kingma2014}.

\subsection{Calculating accelerations from a known DF}
\label{S:Methods:DFtoAcc}

Section~4 of \citetalias{An2021} describes a procedure for calculating gravitational accelerations $\bmath{a}$ given a known DF $f(\bmx, \bmv)$. The calculation is based on an inversion of the CBE under the assumption of dynamical equilibrium. Here, we merely state the main result, i.e.\ the expression giving the acceleration in terms of first derivatives of the DF. At a given point $\bmx$ in configuration space with the cylindrical polar coordinates $(R, \varphi, z)$, the acceleration vector $\bmath{a} = (a_R, a_\varphi, a_z)$ is given by \citepalias[cf.\ eq.~24 in][]{An2021}
\begin{equation}
\label{E:Acc}
     \bmath{a} = \mathbfss{A}^{-1}\sum_{\text{sample}}\bmath{R},
\end{equation}
where $\mathbfss{A} \equiv [\mathsf{A}_{ij}]$ is the matrix with
\begin{gather*}
    \mathsf{A}_{ij} = \sum_{\text{sample}} \frac{\partial f}{\partial\varv_i}\frac{\partial f}{\partial\varv_j};\\
    \bmath{R} = \bmath{\nabla_\varv} f \left(\bmath{\varv \cdot\, \nabla_x} f + \frac{\varv_\varphi^2}{R}\frac{\partial f}{\partial \varv_R} - \frac{\varv_R \varv_\varphi}{R} \frac{\partial f}{\partial \varv_\varphi} \right).
\end{gather*}
The gradient operators here are $\bmath{\nabla_x} \equiv (\partial/\partial R, R^{-1}\partial/\partial\varphi,\partial/\partial z)$ and $\bmath{\nabla_\varv} \equiv (\partial/\partial \varv_R, \partial/\partial\varv_\varphi,\partial/\partial \varv_z)$.

The sums labelled `sample' in these expressions are over a number of suitably chosen points in velocity $(\varv_R, \varv_\varphi, \varv_z)$ space. Formally, we need at least three sample points for the matrix $\mathbfss{A}$ to be invertible. This would suffice if using a DF that is exactly correct, but in our context we work with an approximate reconstructed DF that might be more accurate in some regions of phase space than others. It is therefore safer to increase the sample size. We generally find converged results for $\sim$100 sample points, but to err on the side of caution we choose 1000 sample points every time we calculate an acceleration.

In \citetalias{An2021}, we gave a discussion of how best to sample these velocity points. In particular, we described `zones of avoidance': regions of velocity space worth avoiding. These are typically areas where the DF approaches zero (e.g., near the escape velocity) or where its first derivatives approach zero (e.g., near $\bmv = \bar{\bmv}$). In these areas, small \emph{absolute} errors in the learned DF and its derivatives are large \emph{relative} errors, leading to large errors in the derived accelerations. With these points in mind, we sample velocities throughout the remainder of this paper as follows: writing $\bmv = \bar{\bmv} + \delta\bmv$ and taking $\bar{\bmv}=(0, 220, 0)\,\mathrm{km\,s^{-1}}$, we sample the magnitude of $\delta\bmv$ uniformly between 10 and $50~\mathrm{km\,s^{-1}}$, and its orientation isotropically. We have experimented extensively with more sophisticated sampling schemes but these have not yielded any substantial improvement in accuracy.

Having sampled these velocities, it is then straightforward to evaluate equation~(\ref{E:Acc}) to give the gravitational acceleration at a given spatial location. As noted in \citetalias{An2021}, a particular benefit here of normalizing flows (and their implementation in \textsc{nflows}) is that the learned DF is everywhere exactly differentiable, irrespective of the complexity of the flow architecture. Using automatic differentiation, we can efficiently calculate the exact partial derivatives of the DF without resorting to potentially noisy finite difference schemes.

\begin{table}
	\centering
	\caption{Parameters of six mono-abundance populations (MAPs) in our disc. The metallicity [Fe/H] and abundances [$\alpha$/Fe] are not used further, but simply serve to indicate the type of stellar population being emulated by each MAP. The weights $w_i$ give the relative size of each sub-population, cf.\ eq.~(\ref{E:MAPSuperposition}). Finally, $h_R$ and $\sigma_R$ are two of the five qDF parameters, respectively representing the radial scale lengths and radial velocity dispersions. The remaining three qDF parameters are either held fixed or depend on the listed parameters; see discussion in the text.}
	\label{T:MAPs}
	\begin{tabular}{cccccc} 
		\hline
        $i$ & [Fe/H] & [$\alpha$/Fe] & $w_i$ & $h_R$ & $\sigma_R$ \\
            &        &               &       & kpc   & km\,s$^{-1}$\\
        \hline
        1   & -0.7   &  0.2          & 0.10  & 2.0   & 60.0       \\
        2   & -0.3   &  0.2          & 0.15  & 2.0   & 52.0       \\
        3   & -0.3   &  0.0          & 0.25  & 2.6   & 52.0       \\
        4   &  0.1   &  0.0          & 0.25  & 2.6   & 44.0       \\
        5   &  0.1   & -0.2          & 0.15  & 3.2   & 44.0       \\
        6   &  0.5   & -0.2          & 0.10  & 3.2   & 36.0       \\
        \hline
	\end{tabular}
\end{table}

\begin{table}
	\centering
	\caption{Parameters and component normalizations for our MW models. The adopted models for the three components (bulge, halo, and disc), and thus the meanings of the quoted parameters, are discussed further in the text. The component normalizations $f$ given in the lower part of the table are defined such that component $x$ contributes fraction $f_x$ to the MW circular velocity at 8~kpc.}
	\label{T:MWModels}
	\begin{tabular}{lc}
		\hline
        Parameter                             & Value  \\
        \hline
        Bulge power-law exponent              & -1.8   \\
        Bulge cut-off radius (kpc)            & 1.9    \\
        Halo scale radius (kpc)               & 16     \\
        Disc scale length (kpc)               & 3      \\
        Disc scale height (pc)                & 280    \\
        \hline
        Component normalization               &        \\
        \hline
        $f_\text{b}$                          & 0.05   \\
        $f_\text{h}$                          & 0.35   \\
        $f_\text{d}$                          & 0.6    \\
        \hline
	\end{tabular}
\end{table}

\section{Mock Data}
\label{S:Data}

In \citetalias{An2021}, we tested our technique on a sample of stars distributed spherically in a Hernquist profile. Here, we turn to a more complex test case in local stellar kinematics, i.e.\ the kinematics of stars in a small, heliocentric region of the MW disc. This section describes how the mock dataset is constructed.

\subsection{Distribution Function}
\label{S:Data:DF}

One model for the DF of Galactic disc populations is the `quasi-isothermal' action-based DF (or qDF) first described by \citet{Binney2010, Binney2012b}. \citet{Ting2013} found that the qDF gives a good description of individual `mono-abundance populations' (MAPs) of stars, i.e.\ populations of stars with similar [Fe/H] and [$\alpha$/Fe] abundances. In particular, the qDF gives density profiles that are radially and vertically near-exponential, and velocity dispersion profiles that are radially near-exponential and vertically near-isothermal. Different MAPs will take different parameter values for the qDF (i.e.\ scale lengths and normalizations), and the overall disc population can then be described as a linear superposition of many qDFs. This idea was then applied to real data by \citet{Bovy2013}, who subdivided some 17\,000 G-dwarf stars from SDSS-SEGUE into 43 individual MAPs, modelled each MAP with a qDF, and subsequently derived measurements for the scale length of the MW disc among other important parameters.

For our mock data, we populate the disc with stars drawn from six distinct MAPs. In other words, we construct a DF from the weighted sum:
\begin{equation}
\label{E:MAPSuperposition}
    f(\bmx, \bmv) = \sum_{i=1}^6 w_i\, f_\mathrm{qDF}(\bmx, \bmv | \theta_i).
\end{equation}
Here, $w_i$ is the relative weight and $\theta_i$ the various qDF parameters for population $i$. The qDF requires specification of five scale parameters: the radial scale length $h_R$, the $R$ and $z$ velocity dispersions in the disc-plane $\sigma_R$, $\sigma_z$, and the radial scale lengths of these dispersions $h_{\sigma_R}$, $h_{\sigma_z}$. We remark that these numbers are merely scale parameters and not physical, measurable quantities describing the system. For example, a qDF with $\sigma_R = 40~\mathrm{km\,s^{-1}}$ will not necessarily generate a stellar population with exactly that radial velocity dispersion.

In choosing our qDF parameters, we start by picking pairs of elemental abundances [Fe/H] and [$\alpha$/Fe] -- one pair per MAP -- to characterize the six MAPs. These abundances are not used further in our investigation, but serve to give an indication of the type of stellar population being emulated by each MAP, and thus inform the choice of qDF parameters. We manually choose abundances to roughly match those of the Hypatia catalogue of stellar abundances in the local neighbourhood \citep{Hinkel2014}. Note that \citet{Bovy2013} confined their analysis to stars at large heights above the disc plane, and so found a comparatively greater proportion of $\alpha$-rich, old thick disc stars than is found in the Solar neighbourhood by Hypatia. The elemental abundances we adopt for each MAP are given in Table~\ref{T:MAPs}. In matching qDF parameters to these abundances, we then emulate the trends observed by \citet{Bovy2013}: Fe-poor, $\alpha$-old populations have short radial scale lengths $h_R$ and large velocity dispersions $\sigma_{R}$, while Fe-rich, $\alpha$-young populations are opposite on both counts. Alongside the abundances, Table~\ref{T:MAPs} gives the assigned $h_R, \sigma_{R}$ values, as well as the relative weights $w_i$ of the six MAPs. The remaining three qDF parameters are fixed following \citet{Bovy2013}: $\sigma_{z} = \sigma_{R}/\sqrt{3}$ and $h_{\sigma_R} = h_{\sigma_z} = 8~\mbox{kpc}$.

We have thus arrived at a DF that can be used to sample a mock stellar population containing a mix of thick and thin disc sub-populations. 

\begin{figure*}
    \centering
    \includegraphics{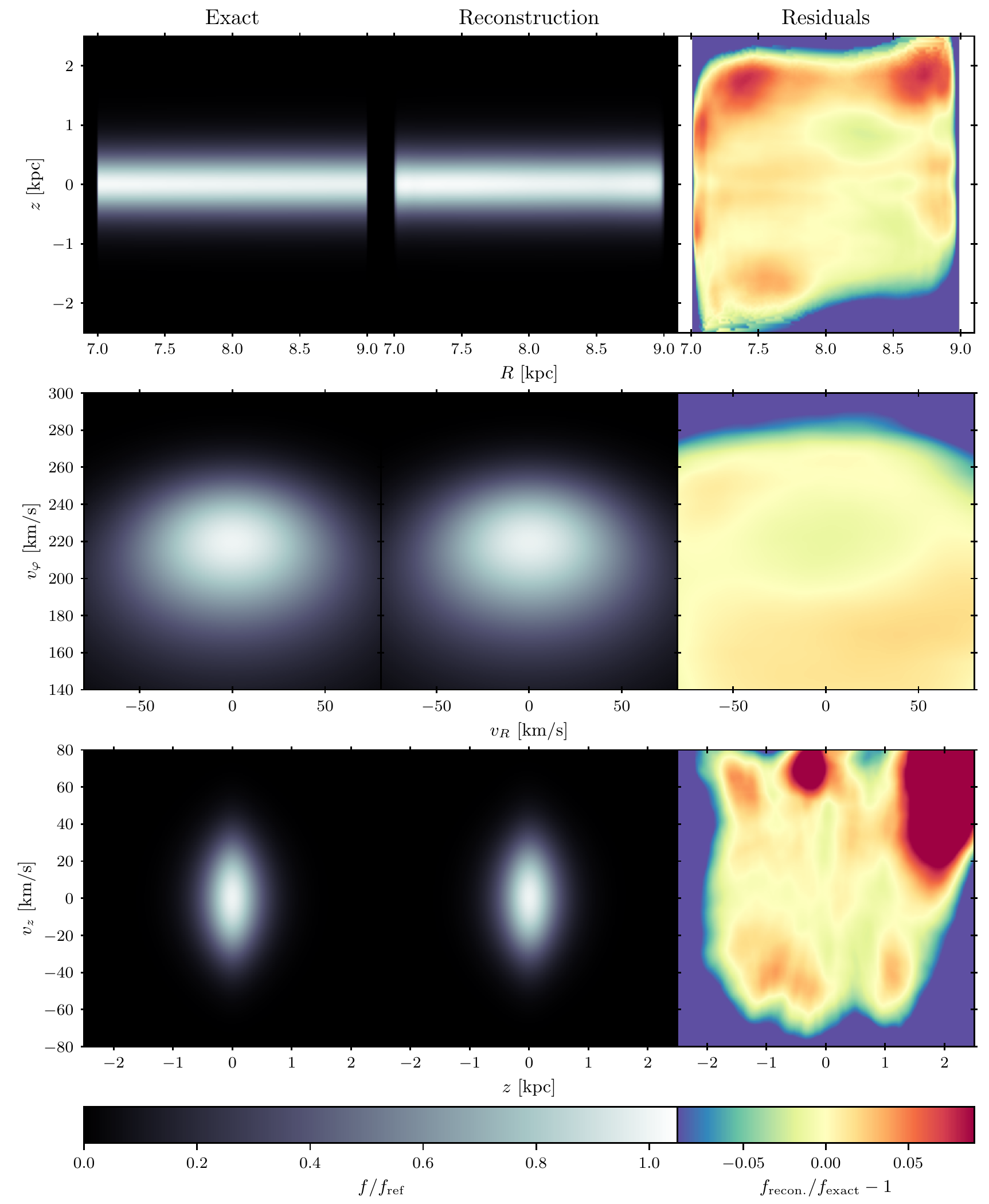}
    \caption{Three slices through phase space: $R$--$z$ \textit{(top)}, $\varv_R$--$\varv_\varphi$ \textit{(middle)}, and $z$--$\varv_z$ \textit{(bottom)}. In each case, the left panel shows the true, underlying DF (eq.~\ref{E:MAPSuperposition}), the middle panel shows the flow-learned DF, and the right panel shows fractional residuals between the two. In each panel, the DF is evaluated by varying two coordinates while holding the remaining coordinates at fixed values. The values of the DF are shown in units of $f_\text{ref}$: the DF evaluated at the solar position ($R=8~\mbox{kpc}, z=10~\mbox{pc}$) and the local standard of rest. This figure demonstrates that our technique is capable of reproducing the underlying stellar DF with excellent accuracy within the well-populated regions of phase space.}
    \label{F:DFs}
\end{figure*}

\begin{figure*}
    \centering
    \includegraphics{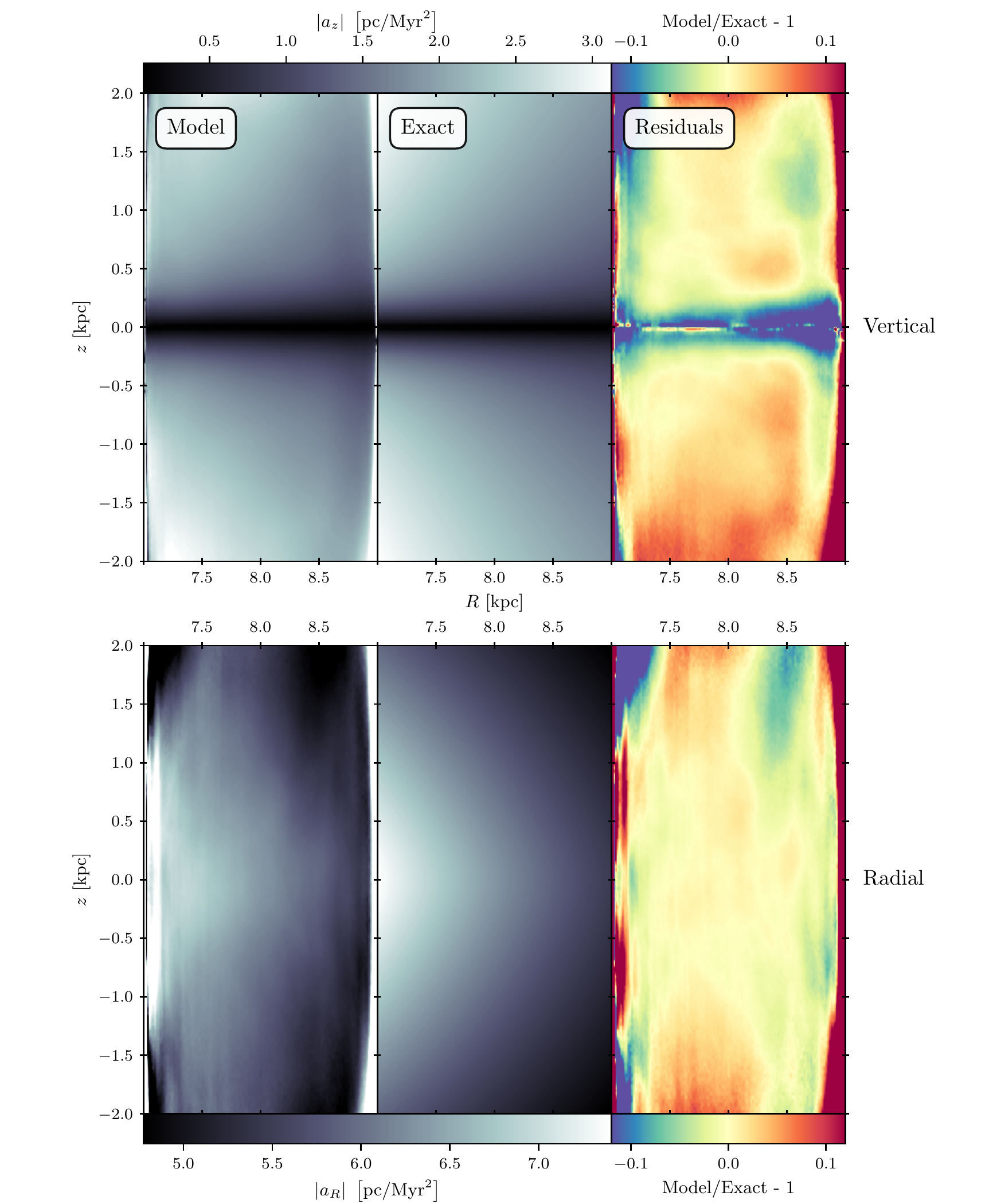}
    \caption{Maps of vertical \textit{(top)} and radial accelerations \textit{(bottom)}. In each case, the left panel plots our reconstructed accelerations, the middle panel the exact accelerations of the model, and the right panel the fractional residuals. This figure demonstrates that, assuming dynamical equilibrium, our technique is capable of calculating accelerations in the solar neighbourhood with excellent accuracy.}
    \label{F:Accs}
\end{figure*}

\subsection{Milky Way Model}
\label{S:Data:MW}

The qDF is a function of orbital actions rather than phase space coordinates, and the conversion from one coordinate system to the other requires the specification of a potential.\footnote{For a given potential, we calculate actions utilizing the the St\"ackel approximation, adopting a focal length of 3.6~kpc \citep{Binney2012a}.} In this way, the underlying gravitational potential is encoded in the stellar kinematics.

The various components and parameters of our adopted MW models are listed in Table~\ref{T:MWModels}. It is identical to the \texttt{MWPotential2014} model of \citet{Bovy2015}: a three-component model, comprising a power-law bulge with exponential cut-off, an NFW dark matter halo, and a Miyamoto-Nagai disc.

\subsection{Sampling}
\label{S:Data:Sampling}

We usually sample $10^6$ stars within an annulus between $R=7$ and 9~kpc, with no restrictions on vertical height $z$. Note that we assume axisymmetry, and so neglect the azimuthal coordinate and sample 5D data, $(R, z, \varv_R, \varv_\varphi, \varv_z)$. Regarding the size of the region, we generally find that for accurate results, survey regions of size $\gtrsim 300~\mbox{pc}$ around the Sun are needed. If smaller regions are used, the flows have difficulty accurately estimating the spatial gradients of the distribution function, leading (via eq.~\ref{E:Acc}) to inaccurate estimates for the acceleration.

The assumption of axisymmetry is not strictly necessary, but leads to a substantial improvement in accuracy. This is due to more than just the reduction in dimensionality: in an axisymmetric or near-axisymmetric system, $\partial f / \partial \varphi$ should be zero or close to zero. However, $\varv_\varphi$ is larger than the other velocity components, due to the Galactic rotation. So, small errors in the estimation of $\partial f / \partial \varphi$ are disproportionately amplified in the $\bmath{\varv \cdot\, \nabla_x}f$ term appearing in equation~(\ref{E:Acc}). Assuming axisymmetry (i.e.\ fixing $\partial f / \partial \varphi=0$) eliminates this effect.

We sample the data directly from the DF (eq.~\ref{E:MAPSuperposition}) using a Markov-Chain Monte-Carlo (MCMC) technique. For this, we use the affine-invariant ensemble sampler implemented in the software package \textsc{emcee} \citep{Foreman2013}. To evaluate the DF in this procedure, we use the qDF implementation and various potential models aboard the software package \textsc{galpy} \citep{Bovy2015}.

\section{Results}
\label{S:Results}

In Section~\ref{S:Results:LocalAccs}, we calculate solar neighbourhood accelerations in our MW model. Then, in Section~\ref{S:Results:Disequilibria}, we consider the effects of disequilibria in the MW disc by perturbing this dataset.

\subsection{The Local Acceleration Field}
\label{S:Results:LocalAccs}

\begin{figure*}
    \centering
    \includegraphics[width = 0.9\linewidth]{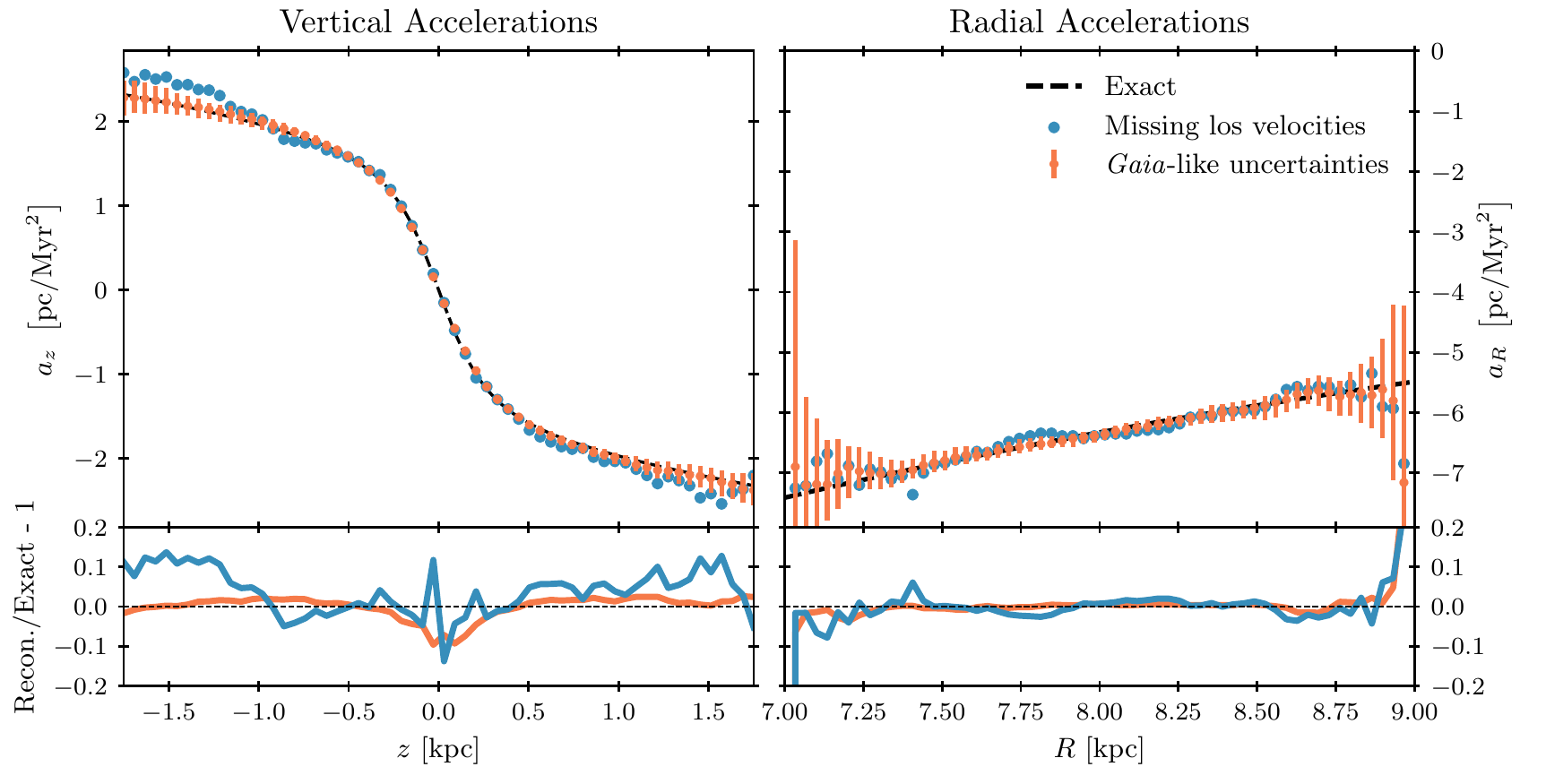}
    \caption{Accelerations measured using a mock dataset with \textit{Gaia}-like \hl{uncertainties in parallax, los velocity, and proper motions (orange points with error bars) and a mock dataset with 50~\% missing los velocities (blue points)}. The vertical accelerations \textit{(left)} are taken at $R=8~\mbox{kpc}$, and the radial accelerations \textit{(right)} are taken at $z=0$. In the case of the mock data with uncertainties, these are propagated by providing a different realization of the error distribution to each flow in the ensemble. The points give the median values across the ensemble, while the error bars show the 16\textsuperscript{th} and 84\textsuperscript{th} percentile values. The black dashed line plots accelerations in the underlying MW model. The smaller panels below show fractional residuals between the measured and true values. Even with realistic errors \hl{or missing los velocities}, our method recovers the underlying acceleration field with sub-percent accuracy in well-populated regions.}
    \label{F:Validation}
\end{figure*}

Training an ensemble of normalizing flows on the mock dataset generated from our MW model, we arrive at a learned DF describing the local population of disc stars. Figure~\ref{F:DFs} depicts this learned DF, alongside the true DF (eq.~\ref{E:MAPSuperposition}) and residuals. Three phase planes are depicted in Figure~\ref{F:DFs}: $R$--$z$, $\varv_R$--$\varv_\varphi$, and $z$--$\varv_z$. In each case, a 2D slice through phase space is shown, i.e.\ two coordinates are varied while the other three are held constant at $R=8~\mbox{kpc}$, $\varv_\varphi=220~\mathrm{km\,s^{-1}}$, $z=\varv_R=\varv_z=0$.

Inspecting the residuals (the right-hand panels of Figure~\ref{F:DFs}), we find excellent percent-level agreement between the true DF and the learned DF in the well-populated region of phase space, i.e.\ within $|\varv - \bar{\varv}| \lesssim 60~\mathrm{km\,s^{-1}}$ and $|z| \lesssim 2~\mbox{kpc}$. The errors only start to grow large at greater velocities or greater heights above or below the mid-plane, where the estimators have very few data points with which to train. 

Another remarkable feature of Figure~\ref{F:DFs}, which might escape notice at first glance, is in the $R$--$z$ plane (top row). Here, there are hard edges in the exact DF at $R=7$ and 9~kpc. These represent the edges of our sample region (Section~\ref{S:Data:Sampling}). When feeding the data to the normalizing flows during the training procedure, the flows are entirely unaware of these hard edges \textit{a priori}. Nonetheless, these sharp edges are detected and reproduced excellently in the model DF, albeit with increased residuals immediately inside the edges. This is a demonstration of the flexibility of normalizing flows, which can deal with sharp transitions in the data.

We now have a DF model that we can input to the machinery of Section~\ref{S:Methods:DFtoAcc} to calculate gravitational accelerations. The result of doing so is shown in Figure~\ref{F:Accs}, which plots derived vertical and radial accelerations alongside the actual accelerations in our MW model. As in Figure~\ref{F:DFs}, the fractional residuals in the well-populated regions are at the sub-percent level. One exception is the region near $z=0$ where the residuals artificially grow large as a result of dividing by small numbers; by eye, it is clear that the agreement remains good in this region. On the other hand, the radial acceleration residuals do truly grow large in the regions immediately near the edges at 7 and 9~kpc, as a result of the DF derivatives being poorly estimated in these regions. As suggested by Figure~\ref{F:DFs}, similar issues also arise at large heights above and below the mid-plane.

Edge effects aside, Figure~\ref{F:Accs} encapsulates the key result of this paper: given $10^6$ stars in equilibrium in an annulus between $R=7$ and 9~kpc, we can calculate the underlying gravitational acceleration field with excellent accuracy.

\begin{figure*}
    \centering
    \includegraphics[width = 0.9\linewidth]{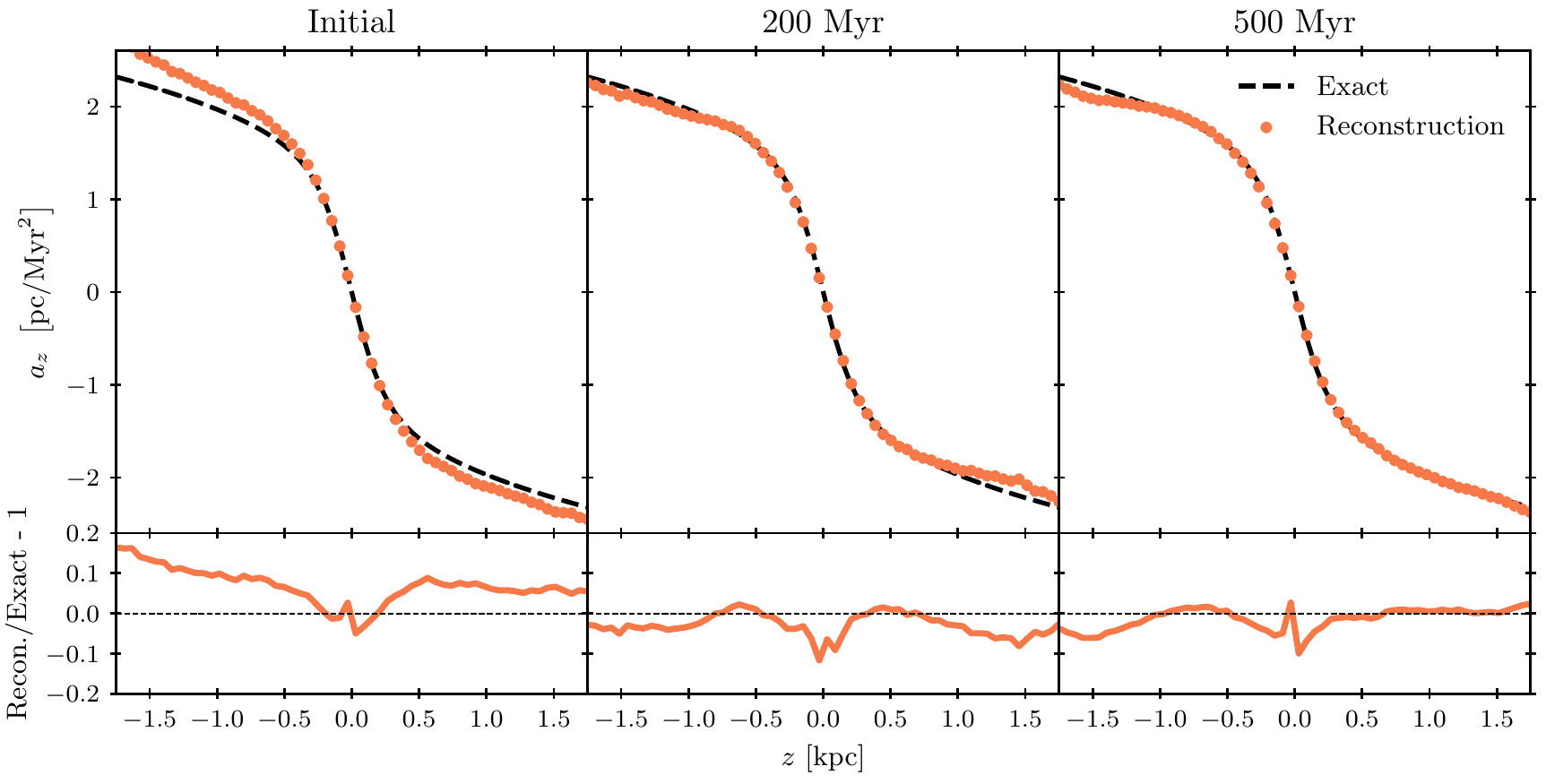}
    \caption{Vertical accelerations derived by applying our methodology to a stellar population following a perturbation. The three main panels present different snapshots in time: immediately following the perturbation (\textit{left}), 200~Myr later (\textit{middle}), and 500~Myr later (\textit{right}). In each case, the coloured points plot the derived accelerations, and the black dashed line plots the true accelerations under our adopted MW model. The three smaller panels below show fractional residuals. This figure illustrates the bias induced by incorrectly assuming equilibrium, and how the bias varies with height and decays with time.}
    \label{F:DiseqAccs}
\end{figure*}

Before forecasting such accuracy for application to the \textit{Gaia} data, it is worth ensuring that this accuracy persists in the presence of realistic errors. We perform this test by adding errors to the \hl{parallax, line-of-sight (los) velocity, and proper motions of each mock star, neglecting errors in the sky positions which we assume to be subdominant. We universally assign Gaussian errors of $\sigma_\varpi = 25~\mbox{$\umu$as}$, $\sigma_\ell = 1~\mbox{km\,s$^{-1}$}$, $\sigma_\wp = 25~\mbox{$\umu$as\,yr$^{-1}$}$ to each star's parallax ($\sigma_\varpi$), los velocity ($\sigma_\ell$), and proper motion ($\sigma_\wp$) respectively, assuming zero covariance. In the real \textit{Gaia} data, these uncertainties correlate with the apparent $G$-band magnitudes: brighter stars have more precise astrometry. Our chosen errors correspond to stars with $G \lesssim 14.5$ in \textit{Gaia} EDR3 (\citealt{Gaia2021}; see also the $\sigma_\wp$ fitting function of \citealt{Dong2021}).} The subset of stars with $G < 14.5$ is the most kinematically robust sample, and at least for \textit{Gaia} DR2, \citet{Schoenrich2019} recommend restricting kinematic analyses to this subset to avoid serious systematic errors. Our assigned errors therefore closely resemble the typical errors in the kind of subset of \textit{Gaia} data to which our method is likely to be applied in future.

We propagate these errors following the method suggested in \citetalias{An2021}: when training an ensemble of flows, each flow is provided with a different dataset, representing a different realization of the error distribution. In practice, each flow takes in the original error-free dataset, transforms the data coordinates to the heliocentric spherical frame, \hl{inverts distances to parallaxes, shifts parallaxes, los velocities and proper motions by random amounts as generated from Gaussian distributions of widths  $\sigma_\varpi$, $\sigma_\ell$, $\sigma_\wp$ respectively,} transforms the data back to the original Galactocentric cylindrical frame, then finally commences the training procedure as normal. Subsequently, the differences in the DFs learned by different flows quantify not only the variability intrinsic to the technique (see Section~\ref{S:Methods:LearningDF}), but also the statistical uncertainty in the training data.

Figure~\ref{F:Validation} plots vertical and radial accelerations measured after propagating uncertainties in this way \hl{(orange points with error bars)}. Each point represents the median measured value across the flow ensemble, while the accompanying error bars give the 16\textsuperscript{th} and 84\textsuperscript{th} percentile values. Reassuringly, the accuracy remains excellent, with sub-percent level residuals everywhere except near the radial edges and large $|z|$ as before.

\hl{Another obstacle facing the application of our technique to real data is that many \textit{Gaia} stars do not have accompanying los velocity data. Of the EDR3 stars with $G < 14.5$, around 30~\% have measured los velocities. The full third data release (scheduled 2022) will fill in many gaps and we can boost the proportion even further by cross-matching the \textit{Gaia} stars with those from independent radial velocity surveys, but it remains inevitable that a significant proportion of the dataset will lack this sixth dimension.}

\hl{There are a number of ways to circumvent this issue. Arguably the simplest is to assume the los velocity selection has minimal kinematic bias, so that all of the stars can be used to learn the stellar density $\nu(\bmx)$, and the subset of stars with available los velocities can be used to learn the (position-dependent) velocity distribution $p(\bmv\, |\, \bmx)$. Normalizing flows can be employed in both cases, and the full distribution function is then given by the product of the two probability distributions. Figure~\ref{F:Validation} shows the results of such an approach, plotting accelerations (blue points) obtained from the same mock data as that used for Figures~\ref{F:DFs}--\ref{F:Accs}, but now with a randomly chosen 50~\% sample of stars taken as having missing los velocity measurement. In other words, the full dataset is used to learn $\nu(R, z)$, but only half of the dataset is used to learn $p(\bmv\, |\, R, z)$. The residuals remain generally small, indicating that the issue of missing los velocities is not insurmountable.}

\hl{There are, however, some issues arising: the residuals are somewhat noisier and grow larger ($\sim 10~\%$) at large $z$. Both of these facts result from half of the data being discarded when learning $p(\bmv\, |\, R, z)$. In particular, the \emph{spatial} gradients of $p(\bmv\, |\, R, z)$ are less well estimated as a consequence. Given that the discarded stars do have two dimensions of velocity information (i.e., their proper motions), we might attain better results by instead retaining these stars and estimating their missing los velocities. A possible technique to do so has been suggested by the work of \citet{Dropulic2021}, which demonstrated that artificial neural networks can be successful in recreating the missing los velocities of \textit{Gaia} stars.
}

\subsection{Disequilibria}
\label{S:Results:Disequilibria}

There is a growing body of evidence for non-equilibrium structure in the stellar kinematics of the MW disc. Whereas the first step of our methodology (learning the DF) assumes only axisymmetry, the second-step (converting to an acceleration field), requires the assumption of dynamical equilibrium so that the time-derivative term can be neglected in the CBE.

In \citetalias{An2021}, we showed that the incorrect assumption of equilibrium leads to a bias in the derived accelerations that is linear in $\partial f / \partial t$. Similarly, \citet{Banik2017} estimated that, under plausible perturbations, the bias induced by incorrect assumption of equilibrium in measurements of vertical accelerations is at the 10~\% level or so. However, their assumed methodology was different from that of the present work, and so the applicability of this estimate is not entirely clear.

Here, we quantify the disequilibrium bias by applying our methodology to a mock dataset representing a perturbed stellar population. To generate this perturbed dataset, we start by employing the method described in Section~\ref{S:Data} to sample an equilibrium dataset comprising $2.5 \times 10^{7}$ stars between $R=1$ and 16~kpc, under our fiducial MW model. Note that this population size gives roughly the desired number of stars ($10^6$) in our region of interest, $7~\mbox{kpc}<R<9~\mbox{kpc}$.

Next, we apply a `kick' to these equilibrium stars, mimicking the procedure of \citet{Li2021}: we randomly choose 10~\% of the stars and boost their vertical velocities by $\delta\varv_z = +20~\mathrm{km\,s^{-1}}$. Such a kick can be understood as being roughly resemblant to the impact of the Sagittarius dwarf passing through the Galactic disc \citep[e.g.,][]{Laporte2019, Bland2021}: under the impulse approximation, $\delta\varv \approx 2 G M / b \varv$, where $M$, $\varv$, and $b$ are respectively the mass, speed, and impact parameter of the perturber. Adopting plausible values of $M=10^{10}~\mbox{M}_\odot$, $\varv=300~\mathrm{km\,s^{-1}}$, and $b=15~\mbox{kpc}$, the resulting kick is $\delta\varv \approx 20~\mathrm{km\,s^{-1}}$.

After applying this perturbation, we evolve the stars' orbits under the (unperturbed) MW potential for 500~Myr, saving snapshots of this evolution at $t=0$, 200, and 500~Myr after the initial perturbation. At each snapshot, we isolate the stars between $R=7$ and 9~kpc and feed them through the pipeline of Section~\ref{S:Methods} to measure accelerations. Figure~\ref{F:DiseqAccs} shows the resulting accelerations at these times.

Immediately after the perturbation, accelerations are everywhere overestimated by 10~\% or so: a similar level of bias to that predicted by \citet{Banik2017}. Note that as in Figure~\ref{F:Accs}, we are still disregarding the residuals immediately around $z=0$. After 200~Myr, the magnitude of the bias has decreased to $\sim 5~\%$, and is confined to larger heights, $|z| \gtrsim 0.5~\mbox{kpc}$. The stars confined to lower heights appear to have equilibriated more quickly, as expected given their shorter dynamical times.

Finally, after 500~Myr, the perturbation appears to have decayed beyond our sensitivity: the residuals are everywhere comparable to the equilibrium case (cf.\ Figure~\ref{F:Accs}). There is a feature in the residuals at $z \approx -1.5~\mbox{kpc}$, but it is unclear whether this is due to lingering effects of the perturbation at large heights or the smaller sampling densities there.

Our finding that the stars have largely equilibriated after 500~Myr is at odds with \citet{Li2021}, who still see a significant bias 500~Myr after an identical perturbation. There are a number of possible causes for this discrepancy. First, a denser Galactic disc has a shorter dynamical time and thus faster equilibriation. However, the difference in the two models doesn't appear to be great enough: the density in our model is only around 35~\% larger, meaning the dynamical time is only around 15~\% shorter (taking $t_\mathrm{dyn} \propto \rho^{-1/2}$). Another possibility is the dimensionality. Under our treatment, we evolve the stellar orbits in three-dimensional configuration space after the initial perturbation, whereas \citet{Li2021} use a one-dimensional approximation. This is tantamount to assuming integrability, as all Hamiltonians with one degree of freedom are exactly integrable. A bundle of trajectories in phase space spreads linearly with time in an integrable system, and so mixing times are longer. A final possibility is the difference in DF models. Whereas we learn a non-parametric DF, \citet{Li2021} fit the stellar kinematics with an analytic DF. It is possible that after 500~Myr, $\partial f/\partial t$ is sufficiently small that equation~(\ref{E:Acc}) can be employed with minimal resulting bias provided the correct DF is used, but the analytic DF used by \citet{Li2021} is not (yet) a good fit for the stars, which still retain some memories of the perturbation in their distribution. In other words, the bias \citet{Li2021} find at 500~Myr might not be directly induced by disequilibrium, but indirectly, via the misapplication of their analytic DF model.

The estimated biases and time-scales shown here can only be used as a rough guide. The real perturbation in the MW disc due to the Sagittarius dwarf could well be larger, and thus induce a longer-lasting bias in the measured accelerations. Moreover, there are other potential sources of \hl{vertical perturbation} beyond the Sagittarius dwarf, such as stellar bar buckling \citep{Khoperskov2019}. \hl{Beyond these vertical perturbations, there are also in-plane perturbations to consider. For example, moving groups (i.e.\ coherent kinematic structures in the local $\varv_x$--$\varv_y$ space) are either dynamical footprints of the Galactic spiral arms and bar~\citep[e.g.,][]{Antoja2008, Michtchenko2018}, or dissolving open clusters and associations~\citep[e.g.,][]{Oh20,Ga21}. These could provide additional contributions to the systematic bias in our estimation of the Galactic acceleration field, but a full accounting is beyond the scope of the present work. However, such effects can be mitigated in practice by masking the stars known to belong to these substructures.}

\begin{figure}
    \centering
    \includegraphics[width = 0.9\linewidth]{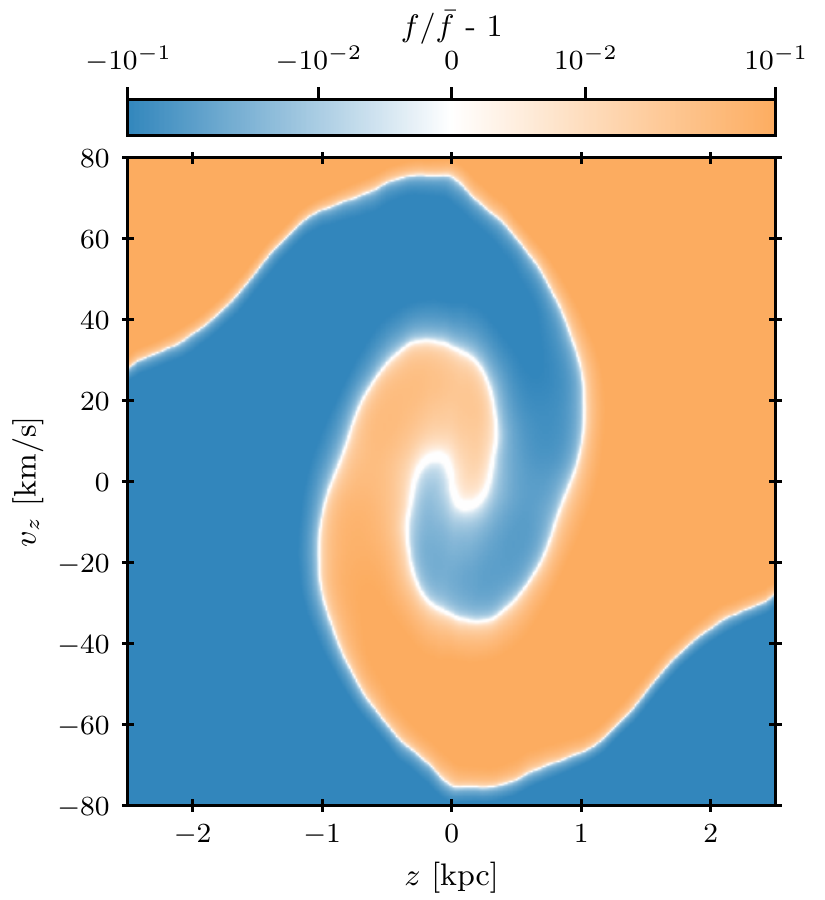}
    \caption{The asymmetric component of the DF trained on the non-equilibrium stellar population, 200~Myr after the perturbation. The asymmetric component is extracted by dividing by a symmetrized DF $\bar{f}$ (definition: eq.~\ref{E:SymmetrisedDF}). A clear phase spiral emerges when plotting the learned DF in this manner.}
    \label{F:Spiral}
\end{figure}

Given this uncertainty, it is worth asking whether our framework provides any way to directly detect the presence of disequilibria. \citet{Li2021} achieved this by comparing their best-fitting model of the DF directly with their perturbed data binned in $z$--$\varv_z$ space, and found that a clear `phase spiral' emerged in the residuals. As a star progresses along its orbit, it exhibits oscillatory motion in the $z$--$\varv_z$ plane. In particular, defining the `vertical energy' $E_z = \varv^2_z/2 + \Phi_z(z)$, where $\Phi_z$ is the vertical part of the galactic potential, stars moves on clockwise `circles' of constant $E_z$. In any potential except a harmonic ($\Phi_z\propto z^2$) potential, the orbital period in this plane is not constant with respect to $E_z$; there is differential rotation. An initial overdensity in $z$--$\varv_z$ space is thus stretched, after the passage of time, into a phase spiral. Eventually, the spiral is stretched and wound to the point where it is no longer detectable, and the population is `phase mixed'. The detection of a phase spiral in a stellar population constitutes clear evidence that the population is not fully phase mixed, i.e.\ not in equilibrium.

Inspired by \citet{Li2021}, we search for a phase spiral in the DF trained on the perturbed data. Unlike in their case, a phase spiral will not emerge in our residuals, because any phase spiral encoded in the data will be similarly encoded in the learned DF. We instead consider a symmetrized DF $\bar{f}$ constructed from the learned DF $f$ via
\begin{equation}
\label{E:SymmetrisedDF}
    \bar{f}(z, \varv_z, R, \varv_R, \varv_\varphi) \equiv \frac{f(z, \varv_z, R, \varv_R, \varv_\varphi) + f(\scalebox{0.75}[1.0]{$-$}z, \scalebox{0.75}[1.0]{$-$}\varv_z, R, \varv_R, \varv_\varphi)}{2}.
\end{equation}
In other words, we take average of $f$ and its 180\degr-rotation in the $z$--$\varv_z$ plane, holding $R, \varv_R, \varv_\varphi$ fixed. We then calculating residuals by comparing $f$ with $\bar{f}$, and so extract any asymmetric component of $f$.

Figure~\ref{F:Spiral} plots $f/\bar{f} - 1$ after 200~Myr, fixing $R=8~\mbox{kpc}$, $\varv_\varphi = 220~\mathrm{km\,s^{-1}}$, $\varv_R = 0$. A remarkably clear phase spiral emerges. The spiral is sharply-defined: the residuals swing from $\sim 10~\%$ to $\sim -10~\%$ very rapidly between the overdense and underdense regions, although the innermost part of the spiral is a bit fainter. We find that the spiral emerges much more cleanly and clearly in these DF slices rather than in projections (i.e.\ integrating $f$ over $\varv_\varphi, \varv_R$), where the residuals are $\sim 1~\%$. This is because the orientation and winding of the spiral vary as a function of $\varv_\varphi$ and $\varv_R$, so that integrating over velocity space serves to partially wash out the signal.

This detection of the phase spiral is an encouraging result: it suggests that a phase spiral would be easy to detect in real data, enabling a straightforward diagnosis of disequilibrium. Moreover, a phase spiral has various uses beyond the simple diagnosis of disequilibrium. In particular, the exact shape of the spiral encodes a wealth of information. For example, \citet{Li2021} fit the phase spiral shape (both their mock spiral and the real Gaia DR2 spiral) to find the time elapsed since the spiral-inducing perturbation. Meanwhile, \citet{Widmark2021SpiralI, Widmark2021SpiralII} use the spiral shape to derive the vertical potential in the Galactic disc, and from there the tightest constraints to date on a thin dark disc.

\section{Comparison with Other Methods}
\label{S:Comparison}

\begin{figure*}
    \centering
    \includegraphics{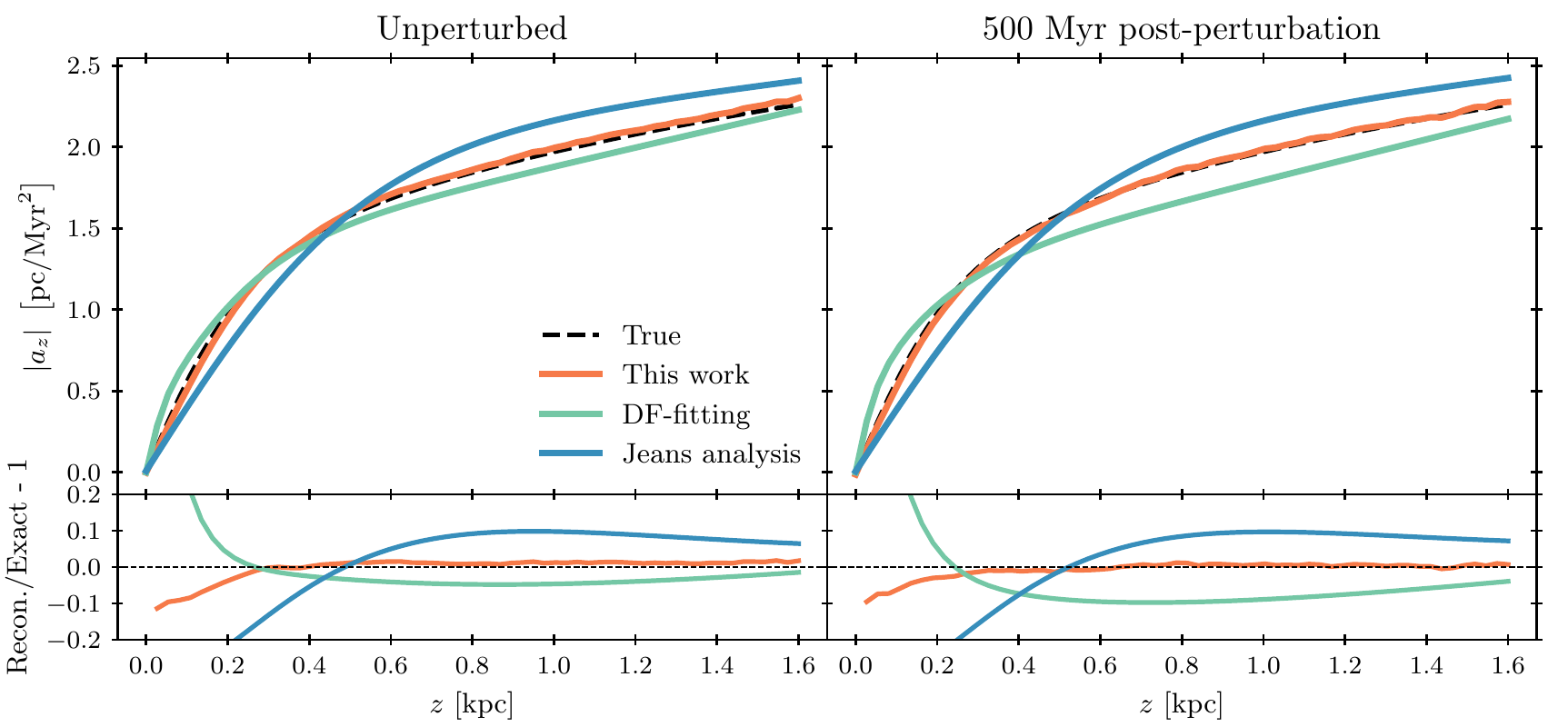}
    \caption{Vertical accelerations at the Solar radius ($R=8~\mathrm{kpc}$) inferred from mock data using different techniques. The two panels correspond to two different mock datasets: the unperturbed dataset of Section~\ref{S:Results:LocalAccs} \textit{(left)}, and the perturbed dataset of Section~\ref{S:Results:Disequilibria}, 500~Myr following the initial perturbation \textit{(right)}. In each case, the black dashed line gives the true accelerations under the assumed MW model, and the coloured lines plot the accelerations inferred via the different techniques as labelled in the legend. By bypassing various limiting assumptions made by other techniques, our method is capable of more accurately measuring accelerations.}
    \label{F:MethodComparison}
\end{figure*}

Here, we test our technique by a direct performance comparison with other, competing methods. Two such techniques are described in the Introduction: Jeans analysis \citep[e.g.,][]{Salomon2020}, and the 1D DF-fitting approach \citep[e.g.,][]{Widmark2021WeighingDisc}. We outline the two methods here, and give more detailed descriptions of our implementations of them in Appendices~\ref{A:Jeans} and \ref{A:DF} respectively.

In the Jeans analysis of \citet{Salomon2020}, stars are binned into radial and vertical bins, and radial and vertical velocity dispersions are computed in each bin, along with the stellar density. Parametrized functional forms are assumed for the spatial variation of the density and dispersions, and these functions are fit to the values obtained from the bins. Given the functional forms and the best-fit parameters, the vertical Jeans equation is solved to give the vertical acceleration. Along with the assumed functional forms, another key assumption concerns the tilt of the local velocity ellipsoid (i.e.\ the covariance between radial and vertical motions), which is assumed to be spherically aligned. This has been shown empirically to be a generally good assumption, except perhaps very close to the disc plane \citep{Everall2019}.

By contrast, the DF-fitting approach of \citet{Widmark2021WeighingDisc} does not bin the data, but directly fits the positions and motions of individual stars. Here, the key assumptions are that the DF is separable, i.e.\ $f(\bmx, \bmv) = f_\perp(z, \varv_z)f_\parallel(x, y, \varv_x, \varv_y)$, and that the `vertical energy' $E_z \equiv \varv_z^2/2 + \Phi(z)$ is an integral of motion. These assumptions, taken together with an assumed parametrization for the potential $\Phi(z)$, give an analytic expression for $f_\perp$ which can be directly fit to the data. The vertical accelerations are then given by first derivative of $\Phi(z)$, taking the best-fit parameters.

Figure~\ref{F:MethodComparison} shows the results of this test for two mock datasets in particular: the unperturbed dataset studied in Section~\ref{S:Results:LocalAccs}, and the perturbed dataset studied in Section~\ref{S:Results:Disequilibria}, 500~Myr after the initial perturbation. In the unperturbed case (left panel), the vertical accelerations calculated with our method reproduce the underlying model with excellent sub-percent level accuracy, as already demonstrated in Section~\ref{S:Results:LocalAccs}. Here, the DF-fitting approach of \citet{Widmark2021WeighingDisc} performs nearly as well, giving residuals of $\lesssim 5~\%$ everywhere. As the data are binned and only the second moments are considered rather than the full shape in a Jeans analysis, much information content is lost. It is then perhaps predictable that the Jeans approach performs the least well of the three. Turning to the perturbed case (right panel), our measured accelerations are of comparable accuracy to the unperturbed case (cf.\ Fig.~\ref{F:DiseqAccs}, right-hand panel). The accuracy of other two methods, however, appears to worsen. The change is most marked in the DF-fitting approach, where the residuals approximately double to the $\sim 10~\%$ level. This finding lends support to our speculation of Section~\ref{S:Results:Disequilibria}: while our technique assumes equilibrium and is thus susceptible to disequilibrium-induced bias, it less susceptible than other techniques.

\section{Conclusions}
\label{S:Conclusions}

A novel procedure for calculating gravitational accelerations from stellar kinematical data was introduced in \citet{An2021}. In this article, we test the methodology in the context of the neighbourhood of the Sun, with a view towards an upcoming paper in which we apply our technique to the \textit{Gaia} data.

The procedure is split into two stages. First, we `learn' the phase-space distribution function (DF) of the data by training normalizing flows \citep{Rezende2015, Kobyzev2020, Papamakarios2021}. In so doing, we construct a data-driven, non-parametric DF, without recourse to any assumptions about the underlying kinematics of the stars, e.g., we do not assume the stellar populations are isothermal or reduce the problem to one dimension. One assumption we do make is axisymmetry, i.e.\ we learn a five dimensional DF in ($R, z, \varv_R, \varv_\varphi, \varv_z$), but this assumption can be easily relaxed if desired. In the second step of the procedure, we convert the learned DF to gravitational accelerations, via an inversion of the collisionless Boltzmann equation. At this stage, we assume that the stars are in dynamical equilibrium.

To test our method, we apply it to a mock dataset resembling a population of Milky Way disc stars in equilibrium. The stars are drawn from multiple `mono-abundance populations', and so represent a mix of stars mirroring the mix of subpopulations in the real Galactic disc. Following \citet{Bovy2013} and \citet{Ting2013}, we construct the mock dataset by assuming that each mono-abundance population can be individually described by a `quasi-isothermal' DF \citep{Binney2010, Binney2012b}, tracing the underlying bulge+halo+disc Milky Way model of \citet{Bovy2015}. We sample $10^6$ stars between 7 and 9~kpc in Galactocentric radius.

Given this mock dataset, we apply our outlined technique, i.e.\ train normalizing flows to learn the DF, then convert the DF to a map of accelerations. We find an excellent sub-percent level match between the measured radial and vertical accelerations and the underlying acceleration field in the adopted Milky Way model. As we will apply our method to data from the {\it Gaia} satellite \citep{Gaia2021}, we check that this excellent accuracy persists even when \hl{realistic errors are added to the data and propagated to the measured accelerations, and when a substantial proportion of line-of-sight velocities are unavailable}. This is the key result of the paper: given the observed positions and motions of a million bright ($G \lesssim 14.5$) stars within a kiloparsec of the Sun, we can robustly determine the underlying gravitational accelerations.

A potential source of systematic bias in our technique is disequilibrium. In converting a learned DF to accelerations, we assume dynamical equilibrium. We test this by employing our technique on a mock dataset following a perturbation emulating the passage of the Sagittarius dwarf through the outer disc. Immediately following the perturbation, we find that accelerations are overestimated by $\sim 10~\%$. This bias decays over time, until it is no longer detectable in the residuals (at least within $|z| \lesssim 1~\mbox{kpc}$) by 500~Myr. Additionally, we find that a remarkably clear and distinct phase spiral can be extracted from the DF trained on the perturbed dataset. Not only can such a phase spiral be used to diagnose disequilibrium, but its shape can also reveal insights into the Milky Way potential and perturbation history \citep[e.g.,][]{Widmark2021SpiralI, Widmark2021SpiralII, Li2021}.

We compare the performance of our method to that of two other widely used methods: solution of the Jeans equations and fitting the vertical (one-dimensional) distribution function to parametrized models. Using the same mock dataset as an input, our method measures accelerations the most accurately. This is particularly true in the aftermath of a perturbation, suggesting that our technique is less susceptible to disequilibrium-induced bias than competing techniques.

In summary, we provide a new algorithm to accurately determine the local acceleration field from stellar kinematical data by non-parametrically reconstructing the stellar DF. We argue that it is the most robust technique yet devised for this purpose. Its strength derives largely from the fact that the DF is constructed directly from the data, thereby bypassing the limiting assumptions and model-sensitivity of our existing methods. In the \textit{Gaia} era, such data-driven techniques have the potential to reveal new insights into fundamental physics and the makeup of our Galactic neighbourhood.


\section*{Acknowledgements}

We thank the anonymous referee for their constructive comments and George Papamakarios for useful discussions. APN and CB are supported by a Research Leadership Award from the Leverhulme Trust. We are grateful for access to the University of Nottingham's Augusta HPC service.

\section*{Data Availability}

The code and mock datasets used in this paper have all been made publicly available at \url{https://github.com/aneeshnaik/LocalFlows}.

\bibliographystyle{mnras}
\bibliography{library}

\begin{thebibliography}{}
\makeatletter
\relax
\def\mn@urlcharsother{\let\do\@makeother \do\$\do\&\do\#\do\^\do\_\do\%\do\~}
\def\mn@doi{\begingroup\mn@urlcharsother \@ifnextchar [ {\mn@doi@}
  {\mn@doi@[]}}
\def\mn@doi@[#1]#2{\def\@tempa{#1}\ifx\@tempa\@empty \href
  {http://dx.doi.org/#2} {doi:#2}\else \href {http://dx.doi.org/#2} {#1}\fi
  \endgroup}
\def\mn@eprint#1#2{\mn@eprint@#1:#2::\@nil}
\def\mn@eprint@arXiv#1{\href {http://arxiv.org/abs/#1} {{\tt arXiv:#1}}}
\def\mn@eprint@dblp#1{\href {http://dblp.uni-trier.de/rec/bibtex/#1.xml}
  {dblp:#1}}
\def\mn@eprint@#1:#2:#3:#4\@nil{\def\@tempa {#1}\def\@tempb {#2}\def\@tempc
  {#3}\ifx \@tempc \@empty \let \@tempc \@tempb \let \@tempb \@tempa \fi \ifx
  \@tempb \@empty \def\@tempb {arXiv}\fi \@ifundefined
  {mn@eprint@\@tempb}{\@tempb:\@tempc}{\expandafter \expandafter \csname
  mn@eprint@\@tempb\endcsname \expandafter{\@tempc}}}

\bibitem[\protect\citeauthoryear{{An}, {Naik}, {Evans}  \& {Burrage}}{{An}
  et~al.}{2021}]{An2021}
{An} J.,  {Naik} A.~P.,  {Evans} N.~W.,   {Burrage} C.,  2021, \mn@doi [\mnras]
  {10.1093/mnras/stab2049}, \href
  {https://ui.adsabs.harvard.edu/abs/2021MNRAS.506.5721A} {506, 5721}

\bibitem[\protect\citeauthoryear{{Antoja}, {Figueras}, {Fern{\'a}ndez}  \&
  {Torra}}{{Antoja} et~al.}{2008}]{Antoja2008}
{Antoja} T.,  {Figueras} F.,  {Fern{\'a}ndez} D.,   {Torra} J.,  2008, \mn@doi
  [\aap] {10.1051/0004-6361:200809519}, \href
  {https://ui.adsabs.harvard.edu/abs/2008A&A...490..135A} {490, 135}

\bibitem[\protect\citeauthoryear{{Antoja} et~al.,}{{Antoja}
  et~al.}{2018}]{Antoja2018}
{Antoja} T.,  et~al., 2018, \mn@doi [\nat] {10.1038/s41586-018-0510-7}, \href
  {https://ui.adsabs.harvard.edu/abs/2018Natur.561..360A} {561, 360}

\bibitem[\protect\citeauthoryear{{Banik}, {Widrow}  \& {Dodelson}}{{Banik}
  et~al.}{2017}]{Banik2017}
{Banik} N.,  {Widrow} L.~M.,   {Dodelson} S.,  2017, \mn@doi [\mnras]
  {10.1093/mnras/stw2603}, \href
  {https://ui.adsabs.harvard.edu/abs/2017MNRAS.464.3775B} {464, 3775}

\bibitem[\protect\citeauthoryear{{Binney}}{{Binney}}{2010}]{Binney2010}
{Binney} J.,  2010, \mn@doi [\mnras] {10.1111/j.1365-2966.2009.15845.x}, \href
  {https://ui.adsabs.harvard.edu/abs/2010MNRAS.401.2318B} {401, 2318}

\bibitem[\protect\citeauthoryear{{Binney}}{{Binney}}{2012a}]{Binney2012a}
{Binney} J.,  2012a, \mn@doi [\mnras] {10.1111/j.1365-2966.2012.21757.x}, \href
  {https://ui.adsabs.harvard.edu/abs/2012MNRAS.426.1324B} {426, 1324}

\bibitem[\protect\citeauthoryear{{Binney}}{{Binney}}{2012b}]{Binney2012b}
{Binney} J.,  2012b, \mn@doi [\mnras] {10.1111/j.1365-2966.2012.21692.x}, \href
  {https://ui.adsabs.harvard.edu/abs/2012MNRAS.426.1328B} {426, 1328}

\bibitem[\protect\citeauthoryear{{Binney} \& {Tremaine}}{{Binney} \&
  {Tremaine}}{2008}]{Binney2008}
{Binney} J.,  {Tremaine} S.,  2008, {Galactic Dynamics, 2nd edn}.
{Princeton Univ.\ Press}, Princeton

\bibitem[\protect\citeauthoryear{{Bland-Hawthorn} \&
  {Tepper-Garc{\'\i}a}}{{Bland-Hawthorn} \&
  {Tepper-Garc{\'\i}a}}{2021}]{Bland2021}
{Bland-Hawthorn} J.,  {Tepper-Garc{\'\i}a} T.,  2021, \mn@doi [\mnras]
  {10.1093/mnras/stab704}, \href
  {https://ui.adsabs.harvard.edu/abs/2021MNRAS.504.3168B} {504, 3168}

\bibitem[\protect\citeauthoryear{{Bovy}}{{Bovy}}{2015}]{Bovy2015}
{Bovy} J.,  2015, \mn@doi [\apjs] {10.1088/0067-0049/216/2/29}, \href
  {https://ui.adsabs.harvard.edu/abs/2015ApJS..216...29B} {216, 29}

\bibitem[\protect\citeauthoryear{{Bovy}}{{Bovy}}{2020}]{Bovy2020}
{Bovy} J.,  2020, arXiv e-prints, \href
  {https://ui.adsabs.harvard.edu/abs/2020arXiv201202169B} {p. arXiv:2012.02169}

\bibitem[\protect\citeauthoryear{{Bovy} \& {Rix}}{{Bovy} \&
  {Rix}}{2013}]{Bovy2013}
{Bovy} J.,  {Rix} H.-W.,  2013, \mn@doi [\apj] {10.1088/0004-637X/779/2/115},
  \href {https://ui.adsabs.harvard.edu/abs/2013ApJ...779..115B} {779, 115}

\bibitem[\protect\citeauthoryear{{Buch}, {Leung}  \& {Fan}}{{Buch}
  et~al.}{2019}]{Buch2019}
{Buch} J.,  {Leung} J. S.~C.,   {Fan} J.,  2019, \mn@doi [\jcap]
  {10.1088/1475-7516/2019/04/026}, \href
  {https://ui.adsabs.harvard.edu/abs/2019JCAP...04..026B} {2019, 026}

\bibitem[\protect\citeauthoryear{{Chakrabarti} et~al.,}{{Chakrabarti}
  et~al.}{2020}]{Chakrabarti2020}
{Chakrabarti} S.,  et~al., 2020, \mn@doi [\apjl] {10.3847/2041-8213/abb9b5},
  \href {https://ui.adsabs.harvard.edu/abs/2020ApJ...902L..28C} {902, L28}

\bibitem[\protect\citeauthoryear{{Chakrabarti}, {Chang}, {Lam}, {Vigeland}  \&
  {Quillen}}{{Chakrabarti} et~al.}{2021}]{Chakrabarti2021}
{Chakrabarti} S.,  {Chang} P.,  {Lam} M.~T.,  {Vigeland} S.~J.,   {Quillen}
  A.~C.,  2021, \mn@doi [\apjl] {10.3847/2041-8213/abd635}, \href
  {https://ui.adsabs.harvard.edu/abs/2021ApJ...907L..26C} {907, L26}

\bibitem[\protect\citeauthoryear{{\VAN{De}{de}{de} Salas} \&
  {Widmark}}{{\VAN{De}{de}{de} Salas} \& {Widmark}}{2021}]{deSalas2020}
{\VAN{De}{de}{de} Salas} P.~F.,  {Widmark} A.,  2021, \mn@doi [Rep.\ Prog.\
  Phys.] {10.1088/1361-6633/ac24e7}, \href
  {https://ui.adsabs.harvard.edu/abs/2020arXiv201211477D} {84, 104901}

\bibitem[\protect\citeauthoryear{{Dong-P{\'a}ez}, {Vasiliev}  \&
  {Evans}}{{Dong-P{\'a}ez} et~al.}{2021}]{Dong2021}
{Dong-P{\'a}ez} C.~A.,  {Vasiliev} E.,   {Evans} N.~W.,  2021, \mn@doi [\mnras]
  {10.1093/mnras/stab3361}, \href
  {https://ui.adsabs.harvard.edu/abs/2021MNRAS.tmp.3055D} {in press,
  arXiv:2110.01060}

\bibitem[\protect\citeauthoryear{{Dropulic}, {Ostdiek}, {Chang}, {Liu}, {Cohen}
   \& {Lisanti}}{{Dropulic} et~al.}{2021}]{Dropulic2021}
{Dropulic} A.,  {Ostdiek} B.,  {Chang} L.~J.,  {Liu} H.,  {Cohen} T.,
  {Lisanti} M.,  2021, \mn@doi [\apjl] {10.3847/2041-8213/ac09ef}, \href
  {https://ui.adsabs.harvard.edu/abs/2021ApJ...915L..14D} {915, L14}

\bibitem[\protect\citeauthoryear{{Durkan}, {Bekasov}, {Murray}  \&
  {Papamakarios}}{{Durkan} et~al.}{2019}]{Durkan2019}
{Durkan} C.,  {Bekasov} A.,  {Murray} I.,   {Papamakarios} G.,  2019, in
  Wallach H.,  Larochelle H.,  Beygelzimer A.,  d\textquotesingle Alch\'{e}-Buc
  F.,  Fox E.,   Garnett R.,  eds,  Advances in Neural Information Processing
  Systems Vol. 32, 33rd Conference on Neural Information Systems (NeurIPS
  2019). Curran Associates, Inc., Vancouver, Canada, pp 7511--7522 (\mn@eprint
  {arXiv} {1906.04032})

\bibitem[\protect\citeauthoryear{{Everall}, {Evans}, {Belokurov}  \&
  {Sch{\"o}nrich}}{{Everall} et~al.}{2019}]{Everall2019}
{Everall} A.,  {Evans} N.~W.,  {Belokurov} V.,   {Sch{\"o}nrich} R.,  2019,
  \mn@doi [\mnras] {10.1093/mnras/stz2217}, \href
  {https://ui.adsabs.harvard.edu/abs/2019MNRAS.489..910E} {489, 910}

\bibitem[\protect\citeauthoryear{{Foreman-Mackey}, {Hogg}, {Lang}  \&
  {Goodman}}{{Foreman-Mackey} et~al.}{2013}]{Foreman2013}
{Foreman-Mackey} D.,  {Hogg} D.~W.,  {Lang} D.,   {Goodman} J.,  2013, \mn@doi
  [\pasp] {10.1086/670067}, \href
  {https://ui.adsabs.harvard.edu/abs/2013PASP..125..306F} {125, 306}

\bibitem[\protect\citeauthoryear{{Gagn{\'e}}, {Faherty}, {Moranta}  \&
  {Popinchalk}}{{Gagn{\'e}} et~al.}{2021}]{Ga21}
{Gagn{\'e}} J.,  {Faherty} J.~K.,  {Moranta} L.,   {Popinchalk} M.,  2021,
  \mn@doi [\apjl] {10.3847/2041-8213/ac0e9a}, \href
  {https://ui.adsabs.harvard.edu/abs/2021ApJ...915L..29G} {915, L29}

\bibitem[\protect\citeauthoryear{{Gaia Collaboration} et~al.,}{{Gaia
  Collaboration} et~al.}{2021}]{Gaia2021}
{Gaia Collaboration} et~al., 2021, \mn@doi [\aap]
  {10.1051/0004-6361/202039657}, \href
  {https://ui.adsabs.harvard.edu/abs/2021A&A...649A...1G} {649, A1}

\bibitem[\protect\citeauthoryear{{Green} \& {Ting}}{{Green} \&
  {Ting}}{2020}]{Green2020}
{Green} G.~M.,  {Ting} Y.-S.,  2020, in Machine Learning \& the Physical
  Sciences, Workshop at the 34th Conference on Neural Information Syetems
  (NeurIPS2020 ML4PS). p.~12 (\mn@eprint {arXiv} {2011.04673})

\bibitem[\protect\citeauthoryear{{Guo}, {Liu}, {Mao}, {Xue}, {Long}  \&
  {Zhang}}{{Guo} et~al.}{2020}]{Guo2020}
{Guo} R.,  {Liu} C.,  {Mao} S.,  {Xue} X.-X.,  {Long} R.~J.,   {Zhang} L.,
  2020, \mn@doi [\mnras] {10.1093/mnras/staa1483}, \href
  {https://ui.adsabs.harvard.edu/abs/2020MNRAS.495.4828G} {495, 4828}

\bibitem[\protect\citeauthoryear{{Hagen} \& {Helmi}}{{Hagen} \&
  {Helmi}}{2018}]{Hagen2018}
{Hagen} J. H.~J.,  {Helmi} A.,  2018, \mn@doi [\aap]
  {10.1051/0004-6361/201832903}, \href
  {https://ui.adsabs.harvard.edu/abs/2018A&A...615A..99H} {615, A99}

\bibitem[\protect\citeauthoryear{{Hinkel}, {Timmes}, {Young}, {Pagano}  \&
  {Turnbull}}{{Hinkel} et~al.}{2014}]{Hinkel2014}
{Hinkel} N.~R.,  {Timmes} F.~X.,  {Young} P.~A.,  {Pagano} M.~D.,   {Turnbull}
  M.~C.,  2014, \mn@doi [\aj] {10.1088/0004-6256/148/3/54}, \href
  {https://ui.adsabs.harvard.edu/abs/2014AJ....148...54H} {148, 54}

\bibitem[\protect\citeauthoryear{{Khoperskov}, {Di Matteo}, {Gerhard}, {Katz},
  {Haywood}, {Combes}, {Berczik}  \& {Gomez}}{{Khoperskov}
  et~al.}{2019}]{Khoperskov2019}
{Khoperskov} S.,  {Di Matteo} P.,  {Gerhard} O.,  {Katz} D.,  {Haywood} M.,
  {Combes} F.,  {Berczik} P.,   {Gomez} A.,  2019, \mn@doi [\aap]
  {10.1051/0004-6361/201834707}, \href
  {https://ui.adsabs.harvard.edu/abs/2019A&A...622L...6K} {622, L6}

\bibitem[\protect\citeauthoryear{{Kingma} \& {Ba}}{{Kingma} \&
  {Ba}}{2015}]{Kingma2014}
{Kingma} D.~P.,  {Ba} J.,  2015, in Bengio Y.,  LeCun Y.,  eds, 3rd
  International Conference on Learning Representations; Conference Track
  Proceedings. {ICLR} 2015.
San Diego, CA, USA, p. poster 9 (\mn@eprint {arXiv} {1412.6980})

\bibitem[\protect\citeauthoryear{{Kobyzev}, {Prince}  \& {Brubaker}}{{Kobyzev}
  et~al.}{2021}]{Kobyzev2020}
{Kobyzev} I.,  {Prince} S. J.~D.,   {Brubaker} M.~A.,  2021, \mn@doi [IEEE
  Trans.\ Pattern Analysis Machine Intelligence] {10.1109/TPAMI.2020.2992934},
  \href {https://ui.adsabs.harvard.edu/abs/2019arXiv190809257K} {43, 3964}

\bibitem[\protect\citeauthoryear{{Laporte}, {Minchev}, {Johnston}  \&
  {G{\'o}mez}}{{Laporte} et~al.}{2019}]{Laporte2019}
{Laporte} C. F.~P.,  {Minchev} I.,  {Johnston} K.~V.,   {G{\'o}mez} F.~A.,
  2019, \mn@doi [\mnras] {10.1093/mnras/stz583}, \href
  {https://ui.adsabs.harvard.edu/abs/2019MNRAS.485.3134L} {485, 3134}

\bibitem[\protect\citeauthoryear{{Li} \& {Widrow}}{{Li} \&
  {Widrow}}{2021}]{Li2021}
{Li} H.,  {Widrow} L.~M.,  2021, \mn@doi [\mnras] {10.1093/mnras/stab574},
  \href {https://ui.adsabs.harvard.edu/abs/2021MNRAS.503.1586L} {503, 1586}

\bibitem[\protect\citeauthoryear{{Loebman} et~al.,}{{Loebman}
  et~al.}{2014}]{Loebman2014}
{Loebman} S.~R.,  et~al., 2014, \mn@doi [\apj] {10.1088/0004-637X/794/2/151},
  \href {https://ui.adsabs.harvard.edu/abs/2014ApJ...794..151L} {794, 151}

\bibitem[\protect\citeauthoryear{{Michtchenko}, {L{\'e}pine}, {Barros}  \&
  {Vieira}}{{Michtchenko} et~al.}{2018}]{Michtchenko2018}
{Michtchenko} T.~A.,  {L{\'e}pine} J.~R.~D.,  {Barros} D.~A.,   {Vieira}
  R.~S.~S.,  2018, \mn@doi [\aap] {10.1051/0004-6361/201833035}, \href
  {https://ui.adsabs.harvard.edu/abs/2018A&A...615A..10M} {615, A10}

\bibitem[\protect\citeauthoryear{{Milgrom}}{{Milgrom}}{1983}]{Milgrom1983}
{Milgrom} M.,  1983, \mn@doi [\apj] {10.1086/161130}, \href
  {https://ui.adsabs.harvard.edu/abs/1983ApJ...270..365M} {270, 365}

\bibitem[\protect\citeauthoryear{{Oh} \& {Evans}}{{Oh} \& {Evans}}{2020}]{Oh20}
{Oh} S.,  {Evans} N.~W.,  2020, \mn@doi [\mnras] {10.1093/mnras/staa2381},
  \href {https://ui.adsabs.harvard.edu/abs/2020MNRAS.498.1920O} {498, 1920}

\bibitem[\protect\citeauthoryear{{Papamakarios}, {Pavlakou}  \&
  {Murray}}{{Papamakarios} et~al.}{2017}]{Papamakarios2017}
{Papamakarios} G.,  {Pavlakou} T.,   {Murray} I.,  2017, in Guyon I.,  Luxburg
  U.~V.,  Bengio S.,  Wallach H.,  Fergus R.,  Vishwanathan S.,   Garnett R.,
  eds,  Advances in Neural Information Processing Systems Vol. 30, 31st
  Conference on Neural Information Systems (NIPS 2017). Curran Associates,
  Inc., Long Beach, CA, USA, pp 2338--2347 (\mn@eprint {arXiv} {1705.07057})

\bibitem[\protect\citeauthoryear{{Papamakarios}, {Nalisnick}, {Rezende},
  {Mohamed}  \& {Lakshminarayanan}}{{Papamakarios}
  et~al.}{2021}]{Papamakarios2021}
{Papamakarios} G.,  {Nalisnick} E.,  {Rezende} D.~J.,  {Mohamed} S.,
  {Lakshminarayanan} B.,  2021, J. Machine Learning Res., \href
  {https://ui.adsabs.harvard.edu/abs/2019arXiv191202762P} {22, 57}

\bibitem[\protect\citeauthoryear{{Read}}{{Read}}{2014}]{Read2014}
{Read} J.~I.,  2014, \mn@doi [J. Phys.\ G: Nucl.\ Part.\ Phys.]
  {10.1088/0954-3899/41/6/063101}, \href
  {https://ui.adsabs.harvard.edu/abs/2014JPhG...41f3101R} {41, 063101}

\bibitem[\protect\citeauthoryear{{Rezende} \& {Mohamed}}{{Rezende} \&
  {Mohamed}}{2015}]{Rezende2015}
{Rezende} D.~J.,  {Mohamed} S.,  2015, in Bach F.,  Blei D.,  eds,  Proceedings
  of Machine Learning Research Vol. 37, Proceedings of the 32nd International
  Conference on Machine Learning. PMLR, Lille, France, pp 1530--1538
  (\mn@eprint {arXiv} {1505.05770})

\bibitem[\protect\citeauthoryear{{Salomon}, {Bienaym{\'e}}, {Reyl{\'e}},
  {Robin}  \& {Famaey}}{{Salomon} et~al.}{2020}]{Salomon2020}
{Salomon} J.-B.,  {Bienaym{\'e}} O.,  {Reyl{\'e}} C.,  {Robin} A.~C.,
  {Famaey} B.,  2020, \mn@doi [\aap] {10.1051/0004-6361/202038535}, \href
  {https://ui.adsabs.harvard.edu/abs/2020A&A...643A..75S} {643, A75}

\bibitem[\protect\citeauthoryear{{Sch{\"o}nrich} \& {Dehnen}}{{Sch{\"o}nrich}
  \& {Dehnen}}{2018}]{Schoenrich2018}
{Sch{\"o}nrich} R.,  {Dehnen} W.,  2018, \mn@doi [\mnras]
  {10.1093/mnras/sty1256}, \href
  {https://ui.adsabs.harvard.edu/abs/2018MNRAS.478.3809S} {478, 3809}

\bibitem[\protect\citeauthoryear{{Sch{\"o}nrich}, {McMillan}  \&
  {Eyer}}{{Sch{\"o}nrich} et~al.}{2019}]{Schoenrich2019}
{Sch{\"o}nrich} R.,  {McMillan} P.,   {Eyer} L.,  2019, \mn@doi [\mnras]
  {10.1093/mnras/stz1451}, \href
  {https://ui.adsabs.harvard.edu/abs/2019MNRAS.487.3568S} {487, 3568}

\bibitem[\protect\citeauthoryear{{Schutz}, {Lin}, {Safdi}  \& {Wu}}{{Schutz}
  et~al.}{2018}]{Schutz2018}
{Schutz} K.,  {Lin} T.,  {Safdi} B.~R.,   {Wu} C.-L.,  2018, \mn@doi [\prl]
  {10.1103/PhysRevLett.121.081101}, \href
  {https://ui.adsabs.harvard.edu/abs/2018PhRvL.121h1101S} {121, 081101}

\bibitem[\protect\citeauthoryear{{Silverwood} \& {Easther}}{{Silverwood} \&
  {Easther}}{2019}]{Silverwood2019}
{Silverwood} H.,  {Easther} R.,  2019, \mn@doi [\pasa] {10.1017/pasa.2019.25},
  \href {https://ui.adsabs.harvard.edu/abs/2019PASA...36...38S} {36, e038}

\bibitem[\protect\citeauthoryear{{Sivertsson}, {Silverwood}, {Read}, {Bertone}
  \& {Steger}}{{Sivertsson} et~al.}{2018}]{Sivertsson2018}
{Sivertsson} S.,  {Silverwood} H.,  {Read} J.~I.,  {Bertone} G.,   {Steger} P.,
   2018, \mn@doi [\mnras] {10.1093/mnras/sty977}, \href
  {https://ui.adsabs.harvard.edu/abs/2018MNRAS.478.1677S} {478, 1677}

\bibitem[\protect\citeauthoryear{{Ting}, {Rix}, {Bovy}  \& {van de Ven}}{{Ting}
  et~al.}{2013}]{Ting2013}
{Ting} Y.-S.,  {Rix} H.-W.,  {Bovy} J.,   {van de Ven} G.,  2013, \mn@doi
  [\mnras] {10.1093/mnras/stt1053}, \href
  {https://ui.adsabs.harvard.edu/abs/2013MNRAS.434..652T} {434, 652}

\bibitem[\protect\citeauthoryear{{Widmark}}{{Widmark}}{2019}]{Widmark2019b}
{Widmark} A.,  2019, \mn@doi [\aap] {10.1051/0004-6361/201834718}, \href
  {https://ui.adsabs.harvard.edu/abs/2019A&A...623A..30W} {623, A30}

\bibitem[\protect\citeauthoryear{{Widmark} \& {Monari}}{{Widmark} \&
  {Monari}}{2019}]{Widmark2019a}
{Widmark} A.,  {Monari} G.,  2019, \mn@doi [\mnras] {10.1093/mnras/sty2400},
  \href {https://ui.adsabs.harvard.edu/abs/2019MNRAS.482..262W} {482, 262}

\bibitem[\protect\citeauthoryear{{Widmark}, {de Salas}  \& {Monari}}{{Widmark}
  et~al.}{2021a}]{Widmark2021WeighingDisc}
{Widmark} A.,  {de Salas} P.~F.,   {Monari} G.,  2021a, \mn@doi [\aap]
  {10.1051/0004-6361/202039852}, \href
  {https://ui.adsabs.harvard.edu/abs/2021A&A...646A..67W} {646, A67}

\bibitem[\protect\citeauthoryear{{Widmark}, {Laporte}  \& {de Salas}}{{Widmark}
  et~al.}{2021b}]{Widmark2021SpiralI}
{Widmark} A.,  {Laporte} C.,   {de Salas} P.~F.,  2021b, \mn@doi [\aap]
  {10.1051/0004-6361/202140650}, \href
  {https://ui.adsabs.harvard.edu/abs/2021A&A...650A.124W} {650, A124}

\bibitem[\protect\citeauthoryear{{Widmark}, {Laporte}, {de Salas}  \&
  {Monari}}{{Widmark} et~al.}{2021c}]{Widmark2021SpiralII}
{Widmark} A.,  {Laporte} C.~F.~P.,  {de Salas} P.~F.,   {Monari} G.,  2021c,
  \mn@doi [\aap] {10.1051/0004-6361/202141466}, \href
  {https://ui.adsabs.harvard.edu/abs/2021A&A...653A..86W} {653, A86}

\makeatother
\end{thebibliography}

\appendix

\section{Jeans Analysis}
\label{A:Jeans}

In this Appendix, we describe the Jeans analysis used for method comparison in Section~\ref{S:Comparison}. Except where noted, we follow the procedure of \citet{Salomon2020}, who used the method to measure the vertical force and local density of dark matter using red clump stars from \textit{Gaia DR2}.

The three Jeans equations relate stellar velocity dispersions and densities to gravitational accelerations, and can be obtained by integrating the CBE (eq.~\ref{E:CBE}) over the three velocity dimensions \citep[e.g.][]{Binney2008}. Under the assumptions of axisymmetry and steady-state, the time-independent vertical Jeans equation is
\begin{equation}
\label{E:Jeans}
    \frac{\partial}{\partial z}\left(\nu \sigma_z^2\right) + \frac{1}{R}\frac{\partial}{\partial R}\left(R\nu \sigma_{Rz}^2 \right) = - \nu \frac{\partial \Phi}{\partial z},
\end{equation}
where $\nu$ is the stellar density, and $\sigma_z^2, \sigma_{Rz}^2$ are the $z$--$z$ and $R$--$z$ components of the velocity dispersion tensor. All three of these quantities are themselves functions of $R$ and $z$. Assuming that the local velocity ellipsoid is spherically aligned, $\sigma_{Rz}^2$ is given in turn by
\begin{equation}
\label{E:JeansSigRz}
    \sigma_{Rz}^2 = Rz\frac{\sigma_R^2 - \sigma_z^2}{R^2 - z^2},
\end{equation}
where $\sigma_R^2$ is the $R$--$R$ component of the velocity dispersion tensor.

To proceed, we need to measure $\nu$, $\sigma_z^2$, $\sigma_R^2$ and their derivatives in order to solve equations~(\ref{E:Jeans}, \ref{E:JeansSigRz}) for $a_z \equiv -\partial \Phi/\partial z$. We do this by binning the stars in $R$ and $z$, calculating the density and velocity dispersions in each bin, and fitting these with functional forms. 

Radially, we use only three bins of width 0.6~kpc, centred at $R=7.4$, 8, 8.6~kpc. We discard stars beyond 7.1 and 8.9~kpc. Vertically, we use adaptive bin sizes, with smaller bins closer to the disc plane. The bin sizes are chosen so that exactly 400 stars (across all radii) fall into each vertical bin. With $\sim 10^6$ in each mock dataset, this gives a few thousand vertical bins. After the radial binning is additionally imposed, the stellar count in each 2D bin varies, but there is still a statistically sufficient number of stars in each bin. Note that we assume mirror symmetry around the disc plane, and so invert $z$ and $\varv_z$ for each star below the plane and restrict all analysis to positive $z$.

In each 2D bin, we measure the stellar density $\nu$ and the velocity dispersions $\sigma_z^2$, $\sigma_R^2$, and assign Poisson errors ($\propto 1/\sqrt{N}$) to each measurement. Given these data points, we can fit the functional forms
\begin{gather} 
\nu(R,z) = \nu_0 \sech^2\!\left(\frac{z}{h}\right) \exp\left(-\frac{R-R_\odot}{d_\nu}\right);\label{E:JeansDensity}\\ 
\sigma_z^2(R,z) = \sigma_{z,0}^2\exp\left(-\frac{R-R_\odot}{d_{\sigma_z}}\right) + \alpha z; \label{E:JeansSigz}\\
\sigma_R^2(R,z) = \sigma_{R,0}^2\exp\left(-\frac{R-R_\odot}{d_{\sigma_R}}\right) + \beta z.\label{E:JeansSigR}
\end{gather}
In all, there are 9 free parameters: $\nu_0$, $h$, $d_\nu$, $\sigma_{z,0}^2$, $d_{\sigma_z}$, $\alpha$, $\sigma_{R,0}^2$, $d_{\sigma_R}$, and $\beta$. Note these functional forms differ slightly from those adopted by \citet{Salomon2020}; we found that these functions gave better fits to our mock data, and ultimately more accurate acceleration measurements (cf.\ Fig.~\ref{F:MethodComparison}). This is likely just a reflection of the differences between our mock dataset and the red clump sample investigated by \citet{Salomon2020}.

To fit equations~(\ref{E:JeansDensity}--\ref{E:JeansSigR}) to the measured densities and dispersions, we use an MCMC technique to find the parameters which maximize the Gaussian likelihood of \citet{Sivertsson2018}:
\begin{equation}
    \mathcal{L} = \mathcal{L}_\nu \mathcal{L}_{\sigma_z} \mathcal{L}_{\sigma_R},
\end{equation}
where $\mathcal{L}_x$ represents the likelihood in quantity $x$, given by
\begin{equation}
    \mathcal{L}_x = \prod_i \exp\left[-\frac{1}{2}\left(\frac{x_{\mathrm{data}, i} - x_{\mathrm{model}, i}}{x_{\mathrm{error}, i}}\right)^2\right],
\end{equation}
where the product is over the data points (i.e.\ the measured values in each 2D bin).

Given the best-fitting parameters, everything on the left-hand side of equation~(\ref{E:Jeans}) can be evaluated to give the vertical acceleration.

\section{1D DF-Fitting}
\label{A:DF}

This Appendix describes the DF-fitting approach we use for comparison in Section~\ref{S:Comparison}. It follows the procedure of \citet{Widmark2019a, Widmark2019b, Widmark2021WeighingDisc}, who use it to measure the dynamical matter density in the solar neighbourhood using \textit{Gaia} data.

Here the key assumptions are that the DF is separable: $f(\bmx, \bmv) = f_\perp(z, \varv_z) f_\parallel(x, y, \varv_x, \varv_y)$, the vertical energy $E_z \equiv \varv_z^2/2 + \Phi(z)$ is an integral of motion, and the stellar population comprises three kinematically distinct subpopulations with different velocity dispersions. Under these assumptions, the vertical DF $f_\perp$ can be written as
\begin{equation}
\label{E:1DDF}
    f_\perp(z, \varv_z) = \sum_{j=1}^3  \frac{c_j}{(2\pi\sigma_j^2)^{1/2}}\exp\left(-\frac{\varv_z^2 + 2\Phi(z)}{2\sigma_j^2}\right),
\end{equation}
where $c_j$, $\sigma_j$ are the relative weights and velocity dispersions of subpopulation $j$. Note that $\sum c_j = 1$ and $c_j \geq 0$.

We assume the underlying matter density takes the parametrized form
\begin{equation}
    \rho(z) = \rho_1 \sech^2\!\left(\frac{z}{h_1}\right) + \rho_2 \sech^2\!\left(\frac{z}{h_2}\right) + \rho_3 \sech^2\!\left(\frac{z}{h_3}\right) + \rho_4,
\end{equation}
where the heights $h_1$, $h_2$ and $h_3$ are fixed respectively at 40, 100, and 300~pc. These heights are slightly different from those used by \citet{Widmark2021WeighingDisc}, and we find they give slightly better fits to our mock data. This density model corresponds to a potential
\begin{equation}
\label{E:1DPotential}
    \Phi(z) = 4\pi G \sum_{i=1}^3 \rho_i h_i^2 \ln\cosh\left(\frac{z}{h_i}\right) + 2\pi G \rho_4 z^2.
\end{equation}
Equations~(\ref{E:1DDF}, \ref{E:1DPotential}) together specify an analytic DF model that can be fit directly to the data. There are 9 free parameters in the model: four density normalizations $\rho_i$, three velocity dispersions $\sigma_j$, and two population weights $c_j$ (not three: the third is fixed by $\sum c_j = 1$).

To obtain the best-fitting parameters, we maximize the likelihood\!
\begin{equation}
    \mathcal{L} = \prod_i \frac{S(z_i) f_\perp(z_i, \varv_{z,i})}{\iint S(z) f_\perp(z, \varv_z)\, dz\, d\varv_z},
\end{equation}
where the product is over individual stars and $S(z)$ is the spatial selection function. We keep only stars with $|z| < 2~\mathrm{kpc}$, so $S(z) = 1$ if $|z| < 2~\mathrm{kpc}$ and $S(z) = 0$ otherwise. Integrating the denominator over $\varv_z$, this expression simplifies to
\begin{equation}
    \mathcal{L} = \prod_i \frac{S(z_i) f_\perp(z_i, \varv_{z, i})}{\displaystyle{\int_{z_\mathrm{min}}^{z_\mathrm{max}}\! dz\, \sum_j c_j \exp\left(-\frac{\Phi(z)}{\sigma_j^2}\right)}}.
\end{equation}

We use an MCMC technique to find the parameters which maximize the this likelihood. We find the best results are obtained if the radial range of the data is restricted to $[7.8, 8.2]$~kpc. Once the best-fitting parameters have been found, the vertical accelerations are given by the first derivative of equation~(\ref{E:1DPotential}),
\begin{equation}
    a_z(z) \equiv -\frac{\partial \Phi}{\partial z} = -4\pi G \sum_{i=1}^3 \rho_i h_i \tanh\left(\frac{z}{h_i}\right) + 4\pi G \rho_4 z.
\end{equation}

\bsp	
\label{lastpage}
\end{document}